\pdfoutput=1

\documentclass[sigconf,screen]{acmart}

\renewcommand\footnotetextcopyrightpermission[1]{}
\PassOptionsToPackage{table,xcdraw}{xcolor}
\PassOptionsToPackage{hyphens}{url}
\usepackage{amsfonts}
\usepackage{balance}
\usepackage{color}
\usepackage{graphics}
\usepackage{url}
\usepackage{listings}
\usepackage{multirow}
\usepackage{rotating}
\usepackage{xspace}
\usepackage{algorithm}
\usepackage{subfig}
\usepackage{microtype}
\usepackage{multicol}

\definecolor{mGreen}{rgb}{0,0.6,0}
\definecolor{mGray}{rgb}{0.5,0.5,0.5}
\definecolor{mPurple}{rgb}{0.58,0,0.82}
\definecolor{backgroundColour}{rgb}{0.95,0.95,0.92}

\DeclareCaptionFormat{centercap}
  {\colorbox{white}
     {\parbox{\dimexpr\columnwidth-2\fboxsep}{\centering #1#2#3}}}

\lstdefinestyle{CStyle}{
    backgroundcolor=\color{backgroundColour},   
    commentstyle=\color{mGreen},
    keywordstyle=\color{magenta},
    numberstyle=\tiny\color{mGray},
    stringstyle=\color{mPurple},
    basicstyle=\footnotesize,
    breakatwhitespace=false,         
    breaklines=true,                 
    captionpos=b,                    
    keepspaces=true,                 
    numbers=left,                    
    numbersep=5pt,                  
    showspaces=false,                
    showstringspaces=false,
    showtabs=false,                  
    tabsize=2,
    language=C,
    xleftmargin=12pt,
    xrightmargin=1pt
}

\lstdefinestyle{BashStyle}{
    basicstyle=\footnotesize\ttfamily,
    aboveskip=3pt,
    belowskip=3pt,
    columns=fullflexible
}

\newcommand{\bash}[1]{\hspace{0.1em} {\footnotesize\ttfamily #1} \hspace{0.1em}}
\newcommand{\bashs}[1]{\hspace{0.1em} {\small\ttfamily #1} \hspace{0.1em}}
\newcommand{\directory}[1]{{\small\ttfamily #1}}
\newcommand{\filename}[1]{{\small\ttfamily #1}}

\newcommand{\name}{RPU\xspace}
\newcommand{\names}{RPUs\xspace}
\newcommand{\fname}{Rosebud\xspace}
\newcommand{\lb}{LB\xspace}

\begin{document}

\title{\fname: Making FPGA-Accelerated Middlebox\\Development More Pleasant\\\vspace{0.5cm}}

\author{Moein Khazraee}
\affiliation{
  \institution{Massachusetts Institute of Technology}
  \city{Cambridge}
  \state{MA}
  \country{USA}
}

\author{Alex Forencich}
\affiliation{
  \institution{UC San Diego}
  \city{La Jolla}
  \state{CA}
  \country{USA}
}

\author{George C. Papen}
\affiliation{
  \institution{UC San Diego}
  \city{La Jolla}
  \state{CA}
  \country{USA}
}

\author{Alex C. Snoeren}
\affiliation{
  \institution{UC San Diego}
  \city{La Jolla}
  \state{CA}
  \country{USA} \vspace{0.5cm}
}

\author{Aaron Schulman}
\affiliation{
  \institution{UC San Diego}
  \city{La Jolla}
  \state{CA}
  \country{USA} \vspace{0.5cm}
}

\begin{abstract}
\vskip 0.15em
We introduce an approach to designing FPGA-accelerated middleboxes that simplifies development,
debugging, and performance tuning by decoupling the tasks of hardware-accelerator implementation and software-application programming. \fname is a
framework that links hardware accelerators to a high-performance packet
processing pipeline through a standardized hardware/software  interface.  This separation of concerns allows hardware developers to focus on optimizing
custom accelerators while freeing software programmers to reuse, configure,
and debug accelerators in a fashion akin to software libraries.
We show the benefits of the \fname framework by building a firewall based
on a large blacklist and porting the Pigasus IDS pattern-matching
accelerator in less than a month.  Our experiments demonstrate that \fname
delivers high performance, serving $\sim$200~Gbps
of traffic while adding only 0.7--7 microseconds of latency.
\end{abstract}

\date{}
\maketitle
\pagestyle{plain}

\section{Introduction}

FPGAs have become the preferred platform on which to deploy in-network processing---middleboxes in particular---due to their flexibility:
middleboxes perform a wide variety of network
functions, many of which require hardware acceleration to function at
today's line rates.  For example, intrusion detection systems
(IDS) and SD-WAN middleboxes~\cite{ethernity} leverage hardware acceleration for signature
matching~\cite{pa7000,pigasus} and supporting encrypted tunnels, respectively.
Critically, FPGAs---as opposed to ASICs---also make it possible for
vendors to update their middleboxes after deployment. For
example, release notes from Palo Alto Networks indicate they have
patched bugs in their deployed
FPGA-based firewalls~\cite{pa-relnotes}.

Unfortunately, FPGA development is inherently intricate and laborious, and high-data-rate requirements only further complicate matters.
Network link speeds have increased past 100~Gbps while FPGA clock rates remain plateaued, forcing
middlebox developers to use resources more efficiently, which requires
increasingly intricate knowledge of specific FPGA hardware elements to optimize their designs,
including memory architecture, physical layout, and I/O. Additionally, FPGA
development cycles are much longer than software because any change might
break the logic that orchestrates packet processing, and there is much less
visibility for debugging. 
Hence, despite their performance benefits, FPGAs have seen adoption for only a limited set of middlebox applications.

In this paper, we present the \fname framework for FPGA-based middlebox designs that simplifies development and
debugging.
At the heart of our framework is a new hardware abstraction we call
a \emph{Reconfigurable Packet-processing Unit} (RPU).
RPUs are reconfigurable blocks within the FPGA fabric where we can drop in
customized packet processing accelerators.  RPUs are orchestrated in software
by FPGA-resident RISC-V CPUs, as opposed to difficult-to-maintain hardware
logic.
Our key insight is that orchestration
can be offloaded to (relatively) slow RISC-V CPUs yet still deliver overall
line-rate processing.  Additionally, this software-oriented framework
makes it easier to debug and deploy accelerators because it enables
software-driven debugging tools and supports hardware accelerator
updates at runtime.

Realizing the RPU abstraction requires generalized
supporting hardware. Specifically, \fname load-balances packets
across parallel RPUs, implements a custom memory architecture to facilitate
high-speed packet processing, and provides
an interface between RPUs and the host.  We implement \fname on
top of a widely used Xilinx FPGA board and show that its area and
latency overheads are marginal. We demonstrate that \fname can achieve
200~Gbps for popular middlebox applications including intrusion
detection and firewalling. For instance, we used the \fname framework
to port the engine of a state-of-the-art FPGA-based IDS
implementation~\cite{pigasus} in less than 3 weeks, while adding
runtime update capability to the IDS.  Furthermore, we discuss how
minor changes to one of the current FPGA families can further reduce
the overheads, enabling additional slack for the developer and, in turn,
simplify the development process. 

This work makes the following contributions:

\begin{itemize}
  \item The RPU abstraction facilitates software-based development and debugging for FPGA-accelerated middleboxes.
  \item We implement the \fname framework for load balancing middlebox workloads across packet processors, enabling coarse parallelism and no-pause reconfiguration.
  \item  We demonstrate that these development tools can be used to build
  200-Gbps middleboxes much faster than with existing tools by porting the
  hardware accelerators of a state-of-the-art IDS to \fname, and making a
  firewall, in only one month.
\end{itemize}

\noindent \fname is open-source~\cite{artifact} and the source code is available at:
\url{https://github.com/ucsdsysnet/Rosebud}. Appendix~\ref{app:artifact} provides instructions for how to reproduce the experiments in the paper.

\section{Challenges}
\label{sec:background}

FPGAs have evolved to include features that make them attractive for high-speed
middlebox applications, but these features require developers to forgo familiar
software development paradigms for highly optimized FPGA hardware development.
In this section we outline the key challenges developers
face when implementing FPGA middleboxes and briefly describe how \fname
addresses or alleviates them by empowering developers to focus
on the accelerator design for their desired functionality.

\subsection{Tailoring Logic to FPGA Platforms}

Three features allow FPGAs to achieve line rate processing, but also increase
the barrier to entry for developers. First, FPGA vendors 
tailor FPGAs for efficient high-speed
packet processing. FPGAs now contain ASIC-like hardened logic
for 100-Gbps Ethernet and PCI Express PHY/MAC. They
also contain larger (e.g., 8$\times$) memory cells to enable buffering multiple
packets, providing needed slack for complex, long-running packet processing
tasks (e.g., managing different packet sizes)~\cite{xilinx-uram}. However, these
network-specific resources are provided as bare-bones hardware; middlebox
developers need to understand exactly how they work and implement glue
logic from scratch to meet their specific needs.

Second, manufacturers overcome FPGA clock-speed limitations by
boosting fabric capacity (e.g., logic, memory, and I/O) through 
technologies that increase transistor density on FPGA
chipsets~\cite{agesoffpga} and employing multiple, interconnected
FPGA dies.  Middlebox developers can use these additional resources to
increase throughput by parallelizing their implementations.  In order to put available resources to best use, however,
developers must provide hints to the heuristic-based FPGA development toolchain to
help it achieve a feasible physical layout in the FPGA's fabric.

Third, modern FPGAs can reconfigure a portion of their fabric while the rest of the
logic continues to operate, enabling runtime updates. This technique,
known as partial reconfiguration (PR), is well-suited for 
middlebox developers as it allows them to modify packet processing accelerators at runtime without
any downtime.
PR also enables developers to reach a full implementation by crafting a portion
of the design as a static part, and adding on to the rest of the design incrementally.
However, PR regions constrain
the selection of the configurable regions in the fabric. Excluding these
regions from the static part of the design also reduces the flexibility of resource
selection and lowers total utilization.
PR regions further require sophisticated border logic to avoid clock-frequency reduction,
in addition to extra logic to ensure that the system remains stable
during the reconfiguration.   It requires a knowledgeable developer to
minimize these overheads.

\fname hides the complexity of using platform-specific hardware
by integrating them within the \fname runtime. \fname lays out
partially reconfigurable regions based on the middlebox application's needs and
provides an optimized packet-distribution subsystem to fully addresses die-boundary-crossing challenges.

\subsection{Orchestrating Parallel Units}

While the increased real estate of modern FPGAs makes it possible to parallelize hardware accelerators to implement line-rate processing,
it requires careful orchestration of the parallel units.
There are two key challenges: (1) Different tasks take different amounts of
time.  For example parsing a packet's header takes less time than finding a
string in the payload. (2) Different tasks might also have different access
patterns.  For example, header processing uses a random access pattern and
string matching uses streaming access. Designing the control logic to properly
orchestrate the data flow among different accelerators is crucial---and one of
the most fragile parts of the design, as a single cycle error in the control
can cause corrupted data which is hard to track.  Handling corner cases and
exceptions further complicates this control logic. 

Orchestration constraints further complicate updates to accelerators and limits
reuse of third-party accelerators. A change to a single accelerator might break
the orchestration among the accelerators and require multiple, long compile
and debugging cycles to update the orchestration logic. Similarly, reusing
third-party accelerators requires adapting their data flow and capabilities to
match the rest of the design; designers are often better off reimplementing
their own accelerators. Moreover, standardized
hardware interfaces, such as AXI or AXI Stream~\cite{axi}, only provide
connection for the
accelerators; while separate orchestration logic is still required.

\fname enables use of software for fine-grained orchestration
control. Our key insight is that orchestration tasks are relatively
lightweight compared to accelerator operations, so they can be
implemented in software running on wimpy cores. This also makes debugging
orchestration easier as it can be scaffolded using familiar
software-development tools.  

\fname also supports course-grained parallelism by distributing packet processing across multiple
parallel RPUs. Additionally, it provides a generic messaging system to address the
communication needs among these parallel units.

\subsection{Development and Debugging}

FPGA development is slow. FPGA toolchains take hours to produce a bitstream for
the FPGA image, and
they offer limited debugging visibility (e.g., FPGA developers frequently debug their designs by
looking at simulation waveforms); both significant limitations compared to
software-based middleboxes that benefit from x86 programming and debugging
tools.
Some of these costs can be mitigated by simulating FPGA designs
before building them, but simulations run much slower than real time---a few
thousand packets can take on the order of hours to simulate, much slower than the millions of
packets per second processed at operational line rates. Run-time debugging, which often is necessary
because bugs can be missed in simulation, requires implementing extra debugging logic---with additional development iterations to hunt down the
bug---which further prolongs development. 

FPGA hardware development languages (i.e., HLS) have the potential to
bridge the gap between software and hardware development; however, at
its core HLS is just a tool that generates Verilog, so the developer
needs to be fully aware of hardware restrictions and orchestration
among different hardware modules.  Furthermore, HLS designs are
typically less performant than hand-crafted
Verilog~\cite{clicknp}. Also, HLS does not support software-like debugging
tools, such as break points or memory dumping. It should be noted that developers
can still use HLS-generated accelerators inside \fname's packet
processors for specific accelerator implementations, but need not employ it for the entire middlebox implementation.
In \fname, use of software within FPGA at run time significantly simplifies the
debugging process.

\section{Processing Abstraction}
\label{sec:rosebud}

We overcome the challenges presented in the previous section by introducing a new
abstraction for middlebox packet processing accelerated by FPGA hardware, the
Reconfigurable Packet-processing Unit (\name). 
The \name abstraction achieves the following
goals: (1) reduce the required
expertise needed to use networking-specific FPGA resources,
(2) provide a way to incrementally update hardware accelerators and their
orchestration, and (3) provide developers
with visibility into hardware at runtime for debugging.

\begin{figure}
\centering
\includegraphics[width=0.9\linewidth]{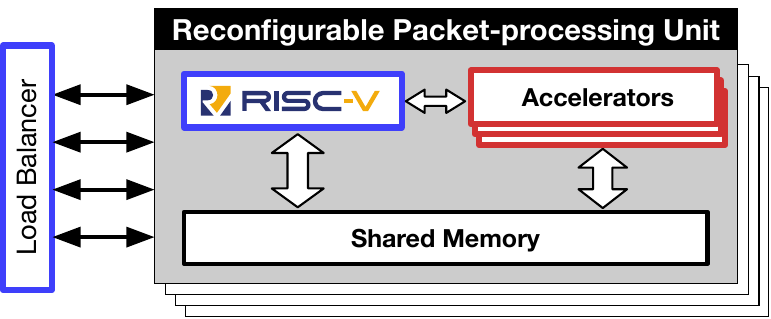}
\caption{
\label{fig:rpu_abstraction}
Overview of the RPU abstraction
}
\end{figure}

\subsection{Reconfigurable Packet-processing Unit} 

An \name consists of a RISC-V core
connected to custom hardware accelerators, all residing inside a partially
reconfigurable FPGA block.
Packets are distributed across the RPUs using a customizable load balancer (\lb).
The software within each \name is in charge of
managing hardware accelerators to process packets; i.e., 
it
orchestrates packet processing by invoking hardware accelerators as it would
call functions in software.
RPUs can be programmed using familiar C-based software abstractions for packet processing,
such as packet descriptors for moving packets using a Direct Memory Access (DMA)
engine.
Figure~\ref{fig:rpu_abstraction} shows an overview of this abstraction.
 Lightweight
processing, such as header parsing, can be implemented in software alone.
To overcome the overhead introduced by the RPU's software-based abstraction and to scale the performance of hardware accelerators, \fname
implements
a customized hardware load balancer
that can be
tailored to match the middlebox's workload.
The user can set the 
policy used by the \lb unit, for example round robin, hash-based,
or even a policy designed specifically for their target middlebox application.

The \name abstraction provides all of the benefits of the 
software development cycle. 
For example, the software running on the RISC-V cores can inject or drop packets
to see if a bug is being caused by them. It also can communicate with a host to provide samples of such packets or implement breakpoint
behavior for certain conditions. Therefore, developers can identify and
mitigate runtime issues by monitoring the state of
accelerators and raising faults that either are handled directly on the software,
or log errors and dump state, if desired on a per-packet basis, for a developer
to inspect. RPUs can even handle doing this on a per-packet basis.
In addition, software running on
the host can dynamically update the hardware accelerators in an RPU at runtime using PR, and handle the
required state transition in software, eliminating the need to design logic
specifically to restore state after reconfigurations.

\subsection{Software-Based Development Environment}
\label{sec:deployment}

\fname enables a developer to only focus on implementing their middlebox
in a single \name before they scale it to run at line-rate. The steps a developer needs to take to go
from an idea to a full middlebox is as follows: (1) Write or choose their
hardware accelerator and connect it within \name.  (2) Write the accompanying
software to support the accelerator and compile it 
alongside the provided \fname libraries. (3) The developer can then test their entire 
software and hardware implementation
in an RPU with a
Python-based simulation framework.
(4) After verifying the desired functionality, they  can then
build their \name with the FPGA development toolchain. The \fname framework's supporting
hardware
will already be placed and routed, including a block available for a custom \lb
module.  (5) Finally, when the FPGA bitstream is
ready, they can use \fname host-side C library and driver to load software
and any constant data (e.g., lookup tables) to the RPU's memory to be shared by the RISC-V cores and accelerators. This process is
further detailed for the case study in Appendix~\ref{app:pig_port}.

For developers, all supporting hardware is pre-made and
pre-laid-out
physically on
the FPGA platform by \fname, allowing them to focus on the
implementation of customized packet-processing software and hardware for their
middlebox. For example, the packet-distribution system crosses
die boundaries in a multi-die FPGA while still meeting high performance and low
latency constraints. The design of the supporting hardware is described in Section~\ref{sec:design}.

\subsection{Flexible Accelerator Orchestration}

{\name}s provide a well-defined interface for software and hardware to
interact: the memory interface between the core and the
accelerators. For instance, software can tell an accelerator what part of a packet to operate on
by passing a pointer---using memory-mapped I/O---to the
data's location in the memory shared with the accelerators.
As long as hardware and software follow the same
memory-based interface, either can be changed without affecting the operation of the
other. 
This effectively brings the commonly used embedded systems programming interface to FPGA middleboxes.
The key insight that makes the RPU abstraction possible is the fact that orchestration takes
much less time than the time it takes to process packets with hardware accelerators. 
However, achieving line-rate with this
this abstraction requires 
a separate high-bandwidth packet
distribution subsystem (Section~\ref{sec:pktdist}).

\subsection{Software-like Debuggability}

An additional benefit of the RPU's shared memory architecture is
the ability to support software-driven hardware debugging.
For runtime debugging, the host can communicate to different components of
\fname. In particular, it has access to memory of each \name to read and
modify the state of accelerators, or even modify how the packets are being processed. The RPU achieves this by allowing
developers to pause and reload software and memory in each \name during
runtime. Furthermore, the framework offers functionality similar to a
breakpoint: RISC-V cores can set status registers that are readable by the host,
and when those status registers change, they RPUs can simultaneously store some
data in host memory, and enter a spin wait for the host to debug the problem.
For example, if the packet distribution part of the \fname framework hangs,
software on the RISC-V can detect the hang using internal timer
interrupt, and send its state to the host.
Furthermore, at runtime the host can send a poke interrupt to an \name
to tell it to stop processing packets, then the host can read the
state of both the accelerators and RISC-V core by dumping the entire RPU
shared memory. 

\fname's architecture also supports
simulating an entire \name's operation, with or without the distribution system,
avoiding the need to lay out a full design and deploy it to FPGA.
We build a Python-based test bench framework based on Cocotb~\cite{cocotb}.
Developers can then link in the hardware accelerators and software they want to
test, and run full simulation of the \name's operation.
Another benefit of using Python is availability of several libraries, such as
Scapy~\cite{scapy} to generate test cases. We also provide a Python
library with the same API available between host and \name.

\section{Hardware Design}
\label{sec:design}

\begin{figure}
\centering
\includegraphics[width=\linewidth]{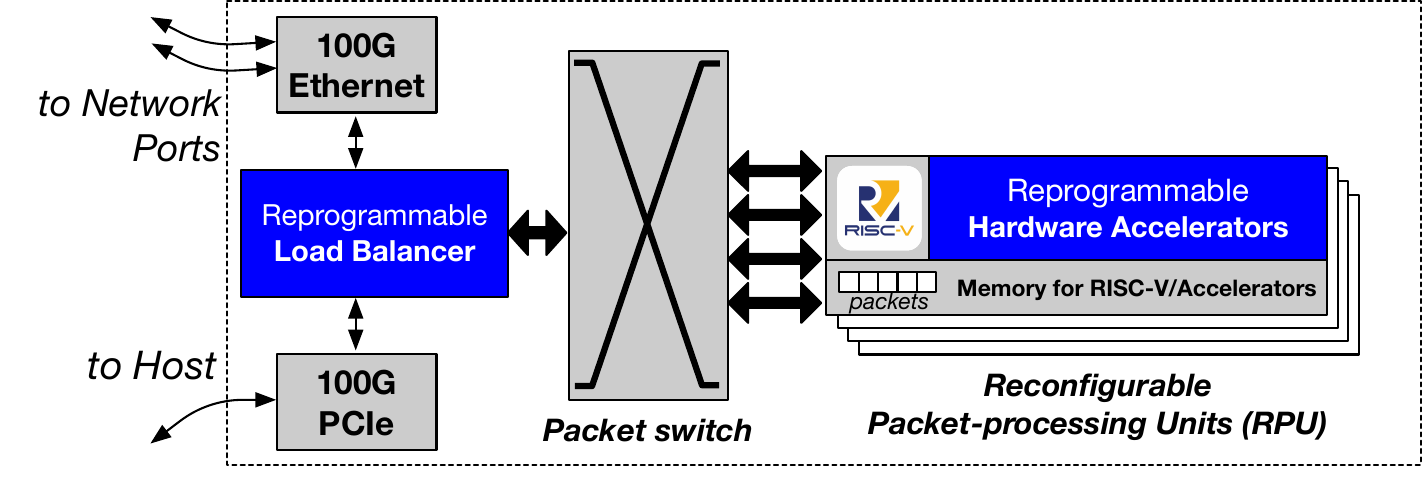}
\caption{
\label{fig:Rosebud_overview}
Overview of the \fname framework
}
\end{figure}

In this section we describe the hardware components in the \fname framework that
enable the RPU abstraction.
Figure~\ref{fig:Rosebud_overview} shows an overview of these components, and
which part of the abstraction they support. 
We start by describing the hardware architecture of the RPU itself, namely the
design of the shared-memory subsystem that enables seamless high-speed
communication between the RISC-V core and the hardware accelerators. Then we
describe the customizable, hardware load balancer that distributes packets across parallel RPUs
to achieve line rates.
Finally, we describe the high-rate packet switch
that distributes packets to and from the external interfaces and host, and across the
RPUs. 

After introducing each component, we briefly overview the interface for the host to
configure that component during runtime, if any.
We also describe how we leverage features of middlebox applications
to implement each component in a performant yet resource-efficient manner,
without a significant decrease in performance or increase in latency over fully
customized FPGA implementations.  
The components described in this section are all
written in Verilog for \fname and come pre-laid out---and placed for the static
portions---in the FPGA logic fabric.
 
\subsection{\name architecture} 
\label{sec:rsu}

Each \name contains two processing components, a RISC-V core and a set of
accelerators.
The RISC-V core and accelerators communicate over two memory-based
interfaces: (1) a basic memory-mapped I/O for configuring and reading
accelerator registers, and (2) a shared-memory subsystem that allows
both the RISC-V core and accelerators to access and modify \emph{packets} that are currently
being processed. This shared memory also holds the \emph{state} of both the
RISC-V core (e.g., stack) and
accelerators (e.g., scratch pad), as well as the \emph{instructions} that the RISC-V
core is executing.

We observe that accelerators and RISC-V cores have different ways of accessing
memory.  Accelerators usually read packet payloads in a streaming manner
(e.g., word-by-word in sequence) to process the entire payload in order, and
perform compute-heavy---i.e., relatively time-consuming---processing. As a result,
accelerators benefit from using larger, higher-latency memories
(e.g., Xilinx's ``Ultra-RAMs'') that can be pipelined to hide the latency while keeping up with line-rate throughput.

In contrast, RISC-V cores have more random accesses; for example parsing a
header and deciding on the next field to read based on the output of that first
read. This random read pattern is inefficient to support with higher-latency
memories. Fortunately, the cores need to access the packet headers which require
less data. As a result, we can copy the packet header and use smaller, lower-latency
memories (e.g., BRAMs) for the RISC-V cores. This contrast provides an
opportunity to design a tailored memory architecture.

\begin{figure}
  \centering
\includegraphics[width=\linewidth]{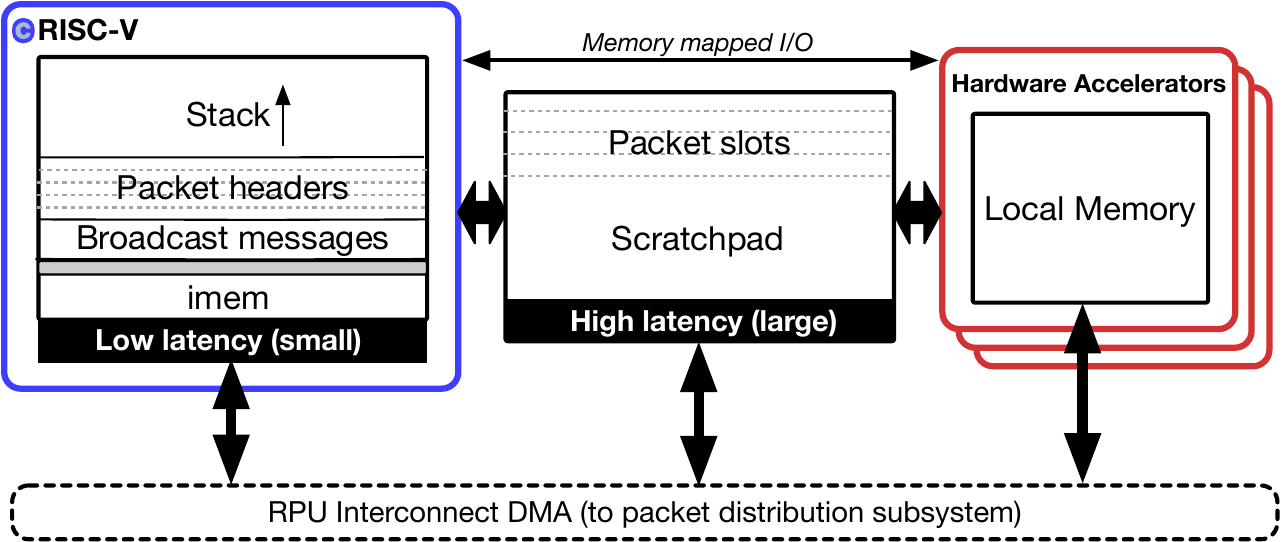}
\caption{
\label{fig:memory_subsystem}
  Memory subsystem in each \name
}
\end{figure}

Figure~\ref{fig:memory_subsystem} shows an overview of the RPU's memory
architecture: we split the memory space into three parts. First, there are
instruction and data memories of the RISC-V core (left) which are small and can be
accessed within a cycle. Then, there is the large packet memory (center), where
the packets arrive at from the packet distribution subsystem. This memory is shared between
the RISC-V and accelerators, and can be used as a scratch pad.
Finally, accelerators can have local memory loaded by the packet distribution subsystem
for lookup tables or similar (right).

An interconnect module provides the interface between the \name and the rest of \fname
that distributes packets to the RPUs (Section~\ref{sec:pktdist}).
This module informs the RISC-V core about arriving packets by giving
it descriptors, and the core can send packets back to the packet distribution
framework by giving descriptors with the desired destination.
The module also provides a control interface from the host to the \name,
for example, it can read and write the status registers, and interrupt
the RISC-V.

The DMA engine inside the interconnect module has access to all these memories. This DMA
engine is customized to copy an incoming packet to the shared packet memory,
and also copy the packet header into the local RISC-V memory to parse the header
with low latency.
DMA engine can also be used to initialize the memories from the
host---before booting the RISC-V core---to load
lookup tables, or read them back for debugging purposes.

One limitation with FPGA memories is that they only have two ports. To avoid
contention, local RISC-V memories have a dedicated port to the core, and the
other port is used for the DMA engine to exchange packet-header data and
facilitate low latency communication among {\name}s (see
Section~\ref{sec:design-msgsubsys}). Since cores sparsely access the packet
memory---e.g., for table look-up once per packet or for change a value in
packets header---we share that port with the DMA engine and give higher
priority to the core. This frees up the other packet memory port for the
accelerators to have exclusive access.
Finally,both ports of the local accelerator memory are dedicated to the
accelerators at runtime, and only during boot or readback---where the
accelerators are not active---the DMA engine can use one of the ports.
For accelerator configuration or result readback, the RISC-V cores uses a
separate memory mapped I/O (MMIO) channel to read/write the accelerators'
registers.

The \name resides entirely inside a PR region,
making it possible to swap it during run time.
The RISC-V cores could have been implemented outside the \name as part
of the supporting architecture. However, due to the need for extra registers on
the border of a reconfigurable region, it would increase the latency of RISC-V
to the shared memories and accelerators. Moreover, by placing the RISC-V
within the \name, it leaves the option open for the developer to customize the 
core and tailor the core capabilities to their needs.

\paragraph{Interface between the host and the {\name}s} We developed an API
to provide the host-based control of {\name}s (details in
Section~\ref{sec:deployment}). In addition, there is an interface to
use the partial reconfiguration process to load  
hardware in an \name. First the host tells the \lb not to send data to the specific \name,
then it waits for the packets within the \name to drain. Next, 
it boots the RISC-V core in the updated \name, and finally tells the \lb to
resume sending data to the
\name. We measured the time to pause, load the new bit file, and
boot a new \name, and it takes 756~milliseconds on average (across 320 loads).

\subsection{Customizable Packet Load Balancer}
\label{sec:design-packetsubsys}

For middleboxes, there is a clear load balancing opportunity to help 
scale the packet processing performance by using \names in parallel.
Indeed, middleboxes often have load balancers that split the load between
parallel servers. 
Use of a load balancer within an FPGA middlebox design results in (1) less burden of
parallelism on the developer to achieve the desired performance, or in other
words less load per \name, (2) it enables laying out partially reconfigurable
regions once and removing that burden from the developer, and (3)
it enables reconfiguring one of the {\name}s during runtime by configuring the \lb to 
offload traffic to 
other RPUs. Developers can customize the \lb policy
to the application's requirements, for instance one that assigns a new packet to the
least-loaded core. We provide TCL scripts to make faster incremental builds
when replacing the LB in the base FPGA image.

We leverage the packet-based data flow in middleboxes to simplify the developer's
design of the \lb. In particular, the \lb
refers
to packet memory in RPUs by a descriptor (slot number). 
Therefore, \lb is only in charge of the load balancing policy and enforces it by
labeling a packet with target \name and memory slot.
These slots are configured by
the software running on RISC-V during boot, where it allocates some slots for packets and
notifies the central \lb about the number of slots and their maximum size.
Similarly, communication between host DRAM and {\name}s is also packetized,
using a different slot number, i.e., DRAM tag.

After the \lb has assigned a destination core for a packet, and that packet arrives at
an \name, the \name's interconnect (described in Section~\ref{sec:rsu}) notifies the
RISC-V of an active descriptor. When sending out a packet, a RISC-V core has two
options: it can ask the RPU interconnect to send it out directly, or it can tell the \lb
which slot is ready to be sent, and the \lb will automatically send the transmit command to the
interconnect. In both cases, the interconnect notifies the \lb about slot being freed
after it is sent out. 
By splitting the control functionality into a
central part---the \lb---and a distributed part---the interconnect, the slot
abstraction also improves scalability of the system.

\paragraph{Interface between the host and \lb}
There is a read/write channel going from the host to the \lb, with
30~bits of separate address space for writing and reading 32~bit 
words. The user can fully customize this channel to configure and
control their \lb during runtime. For instance, the developer uses this channel to
select which cores are used for incoming traffic and which cores are disabled.
They also can read the number of available slots inside the \lb per \name and other
status registers. These data are helpful to detect freezes and starvation.
Also the developer can use the host channel to prepare the \lb for load of a new \name by flushing the slots
in the \lb.

\subsection{Packet Distribution}  
\label{sec:pktdist}

\begin{figure*}
  \centering
  \subfloat[Data distribution flow \label{fig:data_channel}]{%
    \includegraphics[width=.77\textwidth]{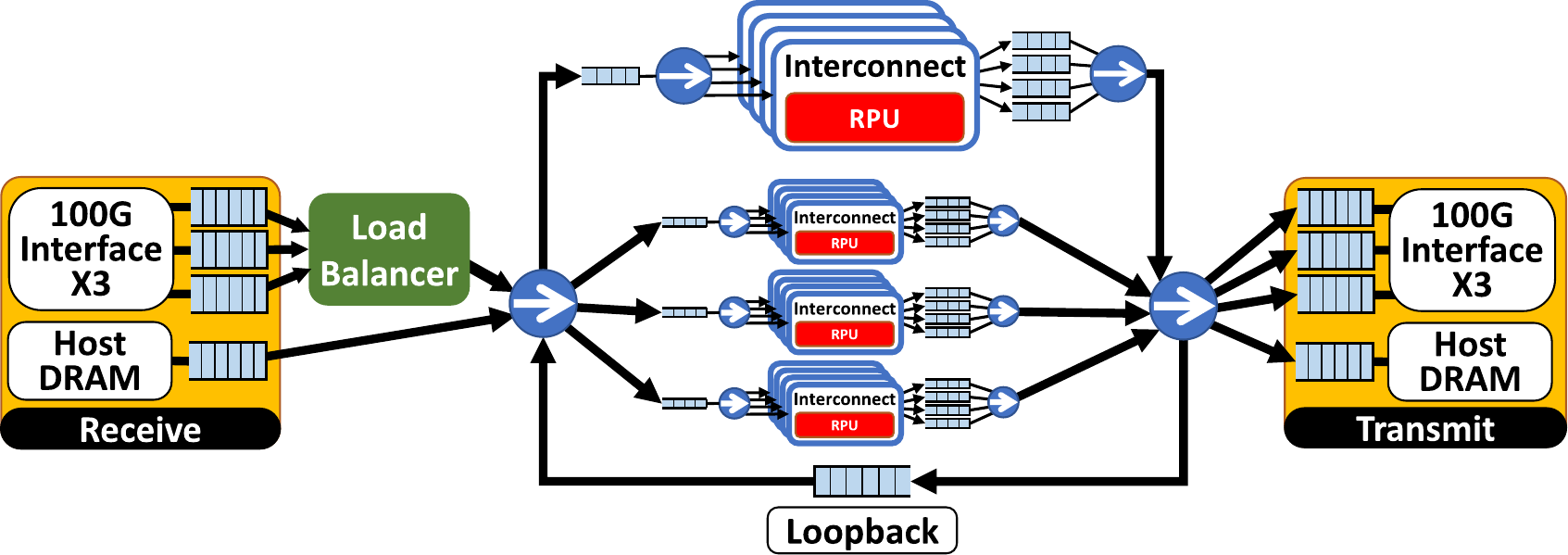}}
  \hspace{2em}
  \subfloat[Control message flow \label{fig:ctrl_channel}]{%
    \includegraphics[width=0.165\textwidth]{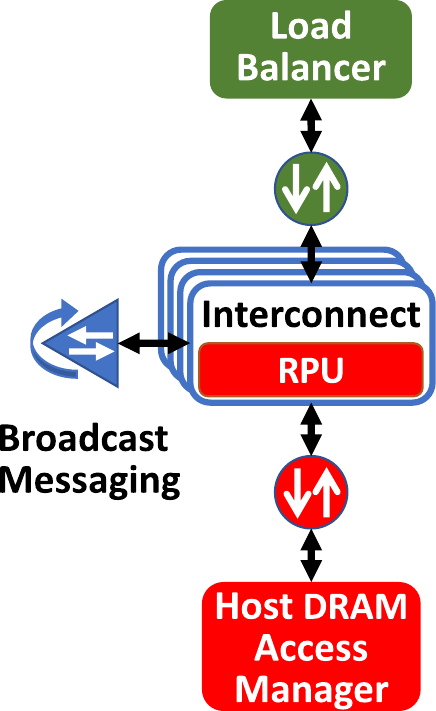}}
  \caption{\label{fig:system}
  \bf Packet distribution and control flow in \fname}
\end{figure*}

Figure~\ref{fig:data_channel} shows an overview of how packets flow from
interfaces to {\name}s. Incoming packets arrive at physical or virtual
Ethernet interfaces, where they are assigned a destination \name by the \lb. There
are two other sources and sinks interfaces for packets: (1) Host DRAM:
used to move data over PCI-e to and from the
host, and (2) loopback: used when an \name wants to send a full packet to
another \name (Section~\ref{sec:design-msgsubsys}). These two interfaces typically
carry much less traffic than network-facing interfaces, so they can share the
same infrastructure without sacrificing throughput.

As shown in Figure~\ref{fig:data_channel}, to make the packet distribution
subsystem more resource efficient, switching is performed in two stages: first
among four \name clusters and then among the {\name}s. This is to achieve the
required performance with as little FPGA fabric resource utilization as possible.  We implement a separate
switch for each RPU cluster that has full throughput incoming, and four links
running at 1/4th throughput outgoing. By using separate FIFOs for
each incoming link inside this switch, the switch achieves non-blocking forwarding: each
FIFO provides bit-width conversion without blocking the other incoming interfaces.
Thus, the only necessary arbitration is when two input interfaces send to the
same \name. We use round-robin-policy arbitration by default; but it can be
replaced with priority policy if desired. Also these switches are unidirectional
and we have separate set of switches for incoming and outgoing traffic, not to
block each other.

We were able to reduce the bit-width on the switches to each \name 
by leveraging the fact that middleboxes do not have strict latency requirements. 
We can already tolerate latency of packets going over the PCIe bus in CPU-based 
middleboxes, as this latency is in the order of
microseconds and it is negligible compared to packets traversing the
Internet, which incurs latency on the order of milliseconds. Therefore, in \fname
we only need to dedicate moderate bandwidth for data communication to each \name 
to reduce the resource requirements of the packet distribution system, while still
incurring less latency than going over the PCIe bus.

Each \name's interconnect is in charge of address handling and
interfacing the packet distribution switch with the RISC-V core, as well as communicating to
the \lb and the host DRAM access manager. This coordination is facilitated by
separate control channels for messages among {\name}s and the \lb, as well as
request messages to host DRAM access manager. These control channels
are separated from the data channels to avoid resource contention among them,
and are shown in Figure~\ref{fig:ctrl_channel}.

\paragraph{Interface between the host and packet distribution subsystem.}
Hosts can read the status counters on all of the
physical and virtual Ethernet interfaces, as well as each \name. These counters
contain the number of transferred bytes, frames, drops, or stalled cycles. They can
shed light to how packets are going through the system, for instance how the \lb is
distributing packets. Therefore, they can reveal to the developer where the
bottlenecks are located.

\subsection{Inter-\name messaging subsystem}
\label{sec:design-msgsubsys}

Middlebox applications
that require stateful packet processing might need to share state between
parallel \names. This can be categorized into two types of message passing:
(1) copying a packet or portion of memory to another \name, or (2) sending
a short message to update the state of the other processors (e.g., table). In a
CPU-based middlebox, both of these can be implemented through cache
coherency, but due to limited memory and the low clock rate on
FPGAs, shared caches are inefficient and impractical to
implement. In \fname, we design tailored messaging systems for these types of
communication to make them more resource efficient.

For sending full packets between \names, we provide a loopback module that can route a
packet between any two \names. The RISC-V cores can ask the \lb 
for a packet slot from a destination \name, and the
packet can be transmitted using the same packet distribution subsystem.
Inter-core packet messaging can also be used to implement a processing chain of heterogeneous
{\name}s with different accelerators and capabilities.

For sending short messages between \names, instead of using a full fledged coherent cache (with
complex eviction and owner core capabilities) \fname features a simplified broadcast messaging
system.  A portion of memory is semi-coherent, where a write to this portion
will eventually be propagated to all the other cores, and they all receive the message
at the exact same time. This broadcast system is shown in Figure~\ref{fig:ctrl_channel}.
This design incurs less contention and overhead than a coherent cache.  We further enable the
use of interrupts to efficiently notify the receiving \name regarding a message.
The RISC-V program can configure the interrupts to be masked based on the target
address. This can be used to send larger messages, where only the last word
causes an interrupt, or to separate data and control messages. \fname further
provides a FIFO for these notifications so as to not lose or reorder them.

\section{Implementation}
\label{sec:impl}

\begin{figure}[t]
\centering
\includegraphics[width=0.97\linewidth]{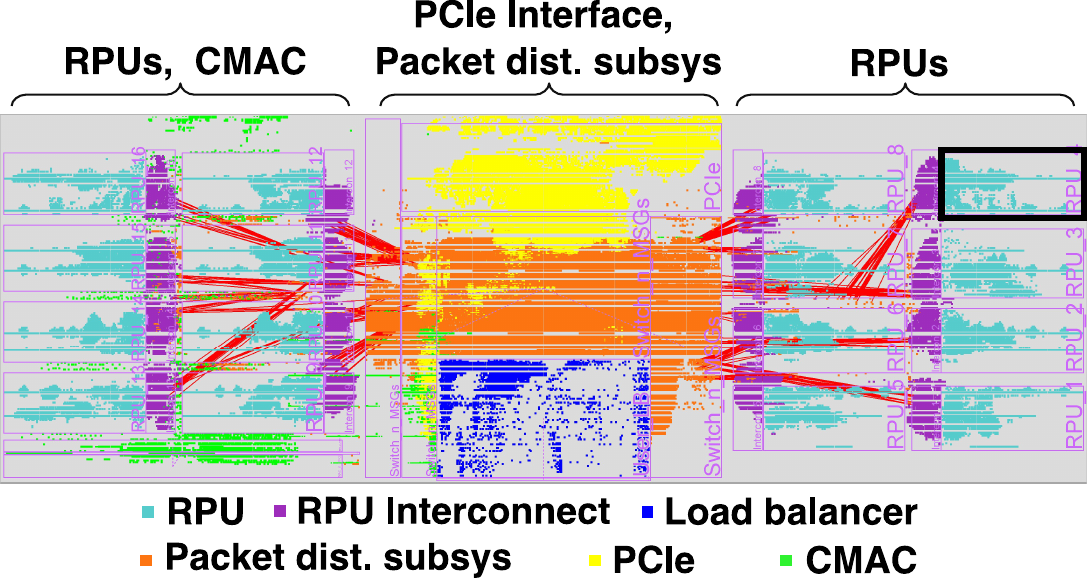}
\vskip -0.6em
\caption{Base FPGA layout with 16 {\name}s}
\label{fig:layout_16G}
\vskip 0.9em
\end{figure}

\begin{figure}[t]
\centering
\includegraphics[width=0.97\linewidth]{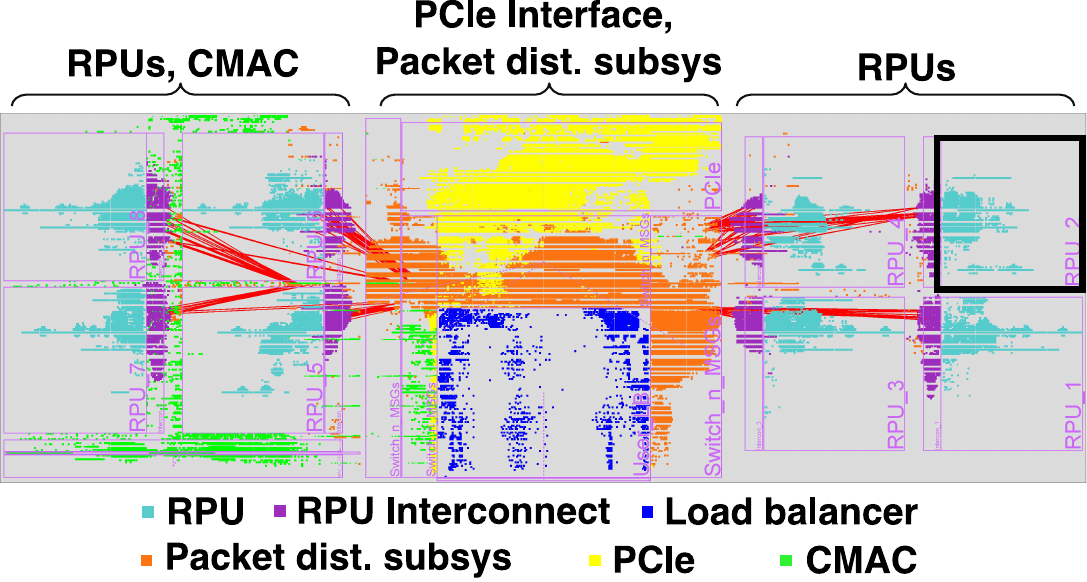}
\vskip -0.6em
\caption{Base FPGA layout with 8 {\name}s}
\label{fig:layout_8G}
\vskip 1.1em
\end{figure}

We implemented \fname on a Xilinx Virtex UltraScale+ FPGA VCU1525 board with an
XCVU9P FPGA chip, shown in Figure~\ref{fig:layout_16G} and
Figure~\ref{fig:layout_8G}. There are 16 and 8 \name versions respectively,
where each \name (cyan) has its own independent PR allocation as well as an
\name interconnect (purple) next to it for
communicating to the rest of the system.  We used the
VexRISC-V~\footnote{\url{https://github.com/SpinalHDL/VexRiscv}} as the core within
each \name. VexRISC-V is a small open source 32-bit RISC-V core with a 5-stage pipeline
that is optimized for FPGAs. It also
can be easily modified to select the developer's desired capabilities. 

\begin{table}[t]
\caption{Base resource utilization for 16 {\name}s}
\vskip -1em
\label{table:utilization_16}
\centering
\resizebox{0.97\linewidth}{!}{%
\setlength{\tabcolsep}{4pt}
\begin{tabular}{|c|c|c|c|c|c|}
  \hline
  Component                 & LUTs            & Registers       & BRAM         & URAM  & DSP \\
  \hline
  Single RPU                &   4541  (0.4\%) &   3788  (0.2\%) &  24  (1.1\%) & 32  (3.3\%) & 0 \\
  Remaining (PR)            &   23298 (2.0\%) &  52132  (2.2\%) &  12  (0.6\%) & 0 & 168 (2.5\%) \\
  \hline
  \lb                       &   8221  (0.7\%) &   22503 (1.0\%) &  0 & 0 & 0 \\
  Remaining                 &   70163 (6.0\%) &  135897 (5.8\%) &  144 (6.7\%) & 48(5.0\%) & 576 (8.4\%)  \\
  \hline
  Single Interconnect       &   2793  (0.2\%) &   2955  (0.1\%) & 0 & 0 & 0  \\
  CMAC                      &   6397  (0.5\%) &  14849  (0.6\%) & 0 &  18 (1.9\%) & 0  \\
  PCIe                      &  41526  (3.5\%) &  63742  (2.7\%) & 110  (5.1\%) & 32(3.3\%) & 0 \\
  Switching                 &  86234  (7.3\%) & 123654  (5.2\%) &  48  (2.2\%) & 64(6.7\%) & 0 \\
  Complete design           & 259713 (22.0\%) & 332636 (14.1\%) & 542 (25.1\%) & 626 (65.2\%) & 0 \\
  \hline
  VU9P device               & 1182240         & 2364480         & 2160         & 960  & 6840         \\
  \hline
\end{tabular}
}
\end{table}

We also allocated another
larger PR block for the \lb (blue). Our current
load balancer implementations are basic and require very few resources, and the
rest of the reserved area is empty for potentially a more sophisticated user \lb.
The main role of partial reconfigurability for the \lb is to isolate its placement and routing,
so if parameter updates to the \lb are not sufficient, a developer can easily replace the
\lb. We do not support partial reconfiguration of \lb during runtime, as it
requires a backup \lb module to switch to for the duration of the
reconfiguration.  The backup \lb adds a significant resource overhead, while for
many middlebox applications updates to the policy can be sufficiently expressed
through passing parameters during runtime.

The physical Ethernet interfaces on the FPGA board are connected via MAC modules
and FIFOs (green). The PCIe modules (yellow) are used for connecting
to the host for control, accessing host DRAM, and providing a virtual
Ethernet interface. Most of these components are provided by the Corundum open-source 100~Gbps
virtual network interface~\cite{corundum}, which 
includes
a driver for the Linux networking stack, enabling \fname's operation as a
NIC. 

Finally, the packet distribution subsystem (orange) is connected to \name
interconnects (red lines show a few of these connections). The widest
switches are 512-bits wide and the narrowest switches are 128-bits
wide, they provide max throughput of 128~Gbps and 32~Gbps respectively. There
is incurred overhead for switch arbitration that reduces their peak performance,
but still the wider switches operate above 100~Gbps. We had to add
several physical constraints to help Vivado (Xilinx's FPGA development
toolchain) in placement of the switches, as
they are wide and span across the FPGA's dies. After
these optimizations, the switching infrastructure uses 54.7\% of the FPGA's die crossing
registers.

Tables~\ref{table:utilization_16} and~\ref{table:utilization_8} show the
utilization breakdown of the 8 and 16 \name \fname runtimes for each of the main components, as
well as the average resource utilization per \name (without any accelerators).
The table also shows the average remaining resources per PR block for each \name, as well as the
remaining resources in the \lb block when using a round robin \lb.
Since the 8 {\name} design
has less arbitration logic than the 16 \name design, we see less resource utilization. 
Across all implementations with all combinations of accelerators, including the
case study accelerators, the maximum routing utilization in vertical or horizontal
direction is 17\%, and typically it is around 12\%. The only hard IP
blocks are the SERDES, PCIe and Gigabit CMAC; the rest are our own open source IP.
We are able to meet timing at 250~MHz
for all designs. 

\begin{table}[t]
\caption{Base resource utilization for 8 {\name}s}
\vskip -1em
\label{table:utilization_8}
\centering
\resizebox{0.97\linewidth}{!}{%
\setlength{\tabcolsep}{4pt}
\begin{tabular}{|c|c|c|c|c|c|}
  \hline
  Component                 & LUTs            & Registers       & BRAM         & URAM  & DSP \\
  \hline
  Single RPU                &   4640  (0.4\%) &   3806  (0.2\%) &  24  (1.1\%) & 32  (3.3\%) & 0 \\
  Remaining (PR)            &   59521 (5.0\%) & 125074  (5.3\%) &  90 (4.2\%) & 32 (3.3\%) & 384 (5.6\%) \\
  \hline
  \lb                       &    7580 (0.6\%) &   22076 (0.9\%) &  0 & 0 & 0 \\
  Remaining                 &  106436 (9.0\%) &  208324 (8.8\%) &  180 (8.3\%) & 96(10.0\%) & 648 (9.5\%)  \\
  \hline
  Single Interconnect       &   2964  (0.3\%) &   3051  (0.1\%) & 0 & 0 & 0  \\
  CMAC                      &   6396  (0.5\%) &  14851  (0.6\%) & 0 &  18 (1.9\%) & 0  \\
  PCIe                      &  41494  (3.5\%) &  63734  (2.7\%) & 110  (5.1\%) & 32(3.3\%) & 0 \\
  Switching                 &  48402  (4.1\%) &  68890  (2.9\%) &  36  (1.7\%) & 32(3.3\%) & 0 \\
  Complete design           & 164699 (13.9\%) & 224404  (9.5\%) & 338 (15.7\%) & 338 (35.2\%) & 0 \\
  \hline
  VU9P device               & 1182240         & 2364480         & 2160         & 960  & 6840         \\
  \hline
\end{tabular}
}
\vskip 0.3em
\end{table}

\section{Evaluation}
\label{sec:eval}

Next we evaluate the performance of the \fname framework. We ran several benchmarks
to understand the limitations of different subsystems within \fname.

\begin{figure*}[t]
  \captionsetup[subfloat]{farskip=-2pt,captionskip=-2pt}
  \centering
  \subfloat[Packet forwarding with 16 {\name}s\label{fig:pktgen_16G}]{%
    \includegraphics[width=0.33\textwidth,trim={1.5em 0 1.5em 0},clip]{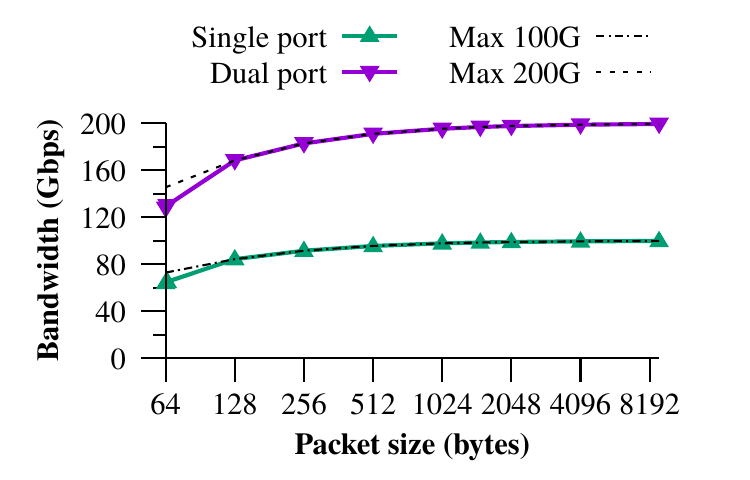}}
  \subfloat[Packet forwarding with 8 {\name}s\label{fig:pktgen_8G}]{%
    \includegraphics[width=0.33\textwidth,trim={1.5em 0 1.5em 0},clip]{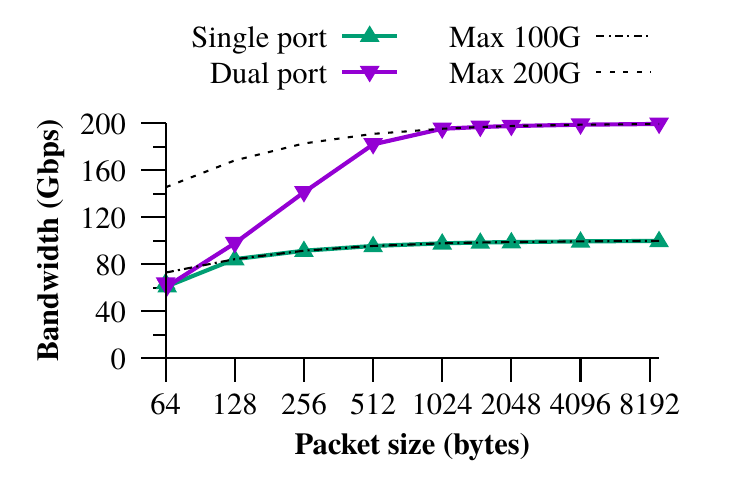}}
  \subfloat[Round-trip latency\label{fig:latency_data}]{%
    \includegraphics[width=0.33\textwidth,trim={2em -1.5em 4em 0},clip]{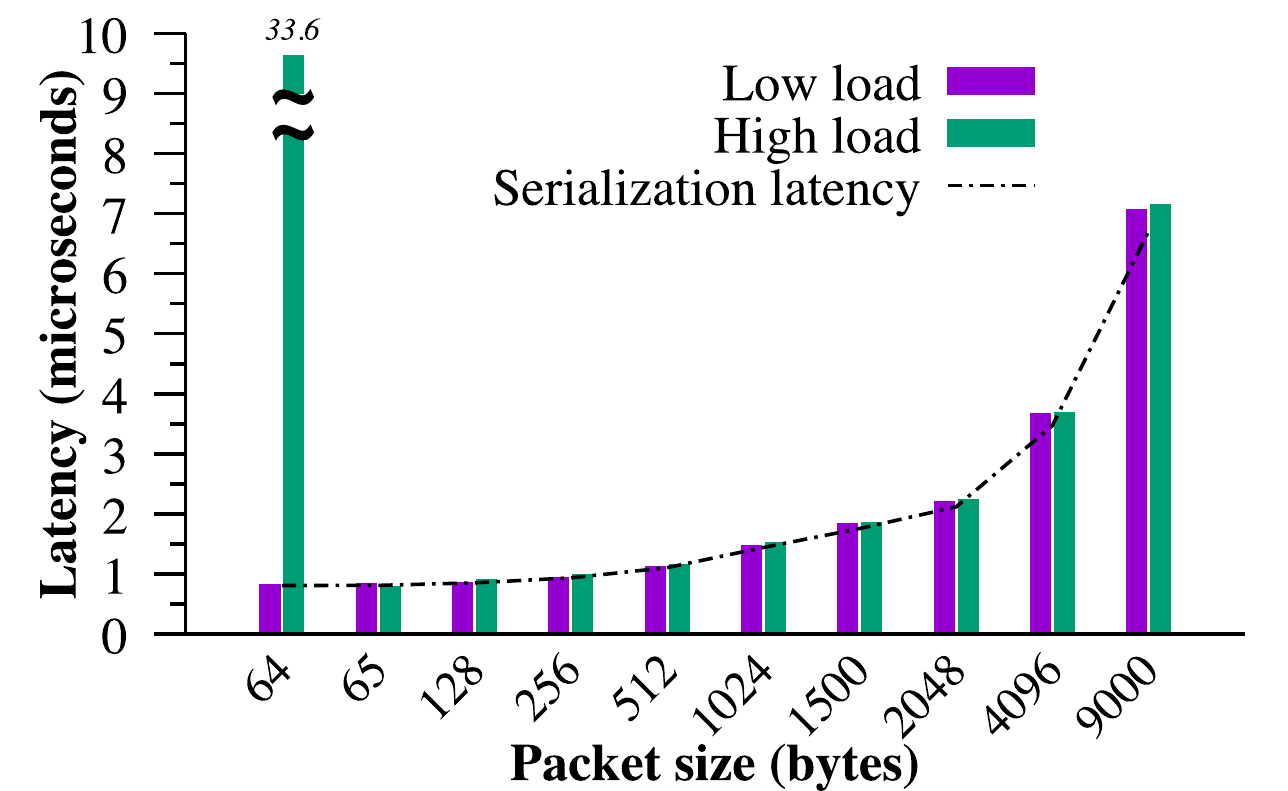}}
\caption{
\label{fig:fwd_eval}
Packet forwarding throughput
with (a) 16, and (b) 8 {\name}s, and 
(c) round-trip latency.}
\end{figure*}

\paragraph{Experiment setup}
Our experiments are conducted using a host with an Intel(R)
Xeon(R) CPU E3-1230 V2 running at 3.3~GHz and a PCIe Gen 3 x16
expansion bus. We installed two separate Xilinx Virtex UltraScale+ FPGA
VCU1525 boards into the PCIe bus, one serves as a traffic source/sink,
while the other runs the system under test.
Both ports of each FPGA are connected to the other FPGA,
each with a 100-Gbps QSFP+ cable, totalling to a bandwidth
of 200~Gbps. We use a round robin
\lb for this evaluation, as well as optimized bare-metal C code on the
RISC-V processors to isolate software overhead from the overhead inherent to the framework.

\subsection{Forwarding Throughput}
\label{sec:fwd_throughput}

We tested the forwarding performance of the framework as a function of
packet size. We consider packet sizes in the range of 64 to 8192 bytes by powers of
two (excluding the 4-byte FCS), including the worst case of 65 bytes and the typical
datacenter MTU packet sizes of 1,500 and 9,000 bytes. We sent packets from the
tester FPGA interfaces and run a simple forwarder program on the \names to observe what portion of the packets are successfully forwarded.
 We use a 16 {\name} design for the packet generation, and it saturates the
maximum line rate of each packet size on the 200~Gbps link, other than for
64-byte packets which achieves 88\% of the maximum rate at 200~Gbps (or 250~MPPS), and for 
65-byte packets which achieves 89\% of the maximum rate at both 100~Gbps and
200~Gbps (or 125~MPPS and 250~MPPS, respectively).

Figure~\ref{fig:fwd_eval} (a) and (b) show the maximum forwarding rate as a
function of packet size for 16 and 8 {\name}s respectively. The maximum
theoretical effective rate is depicted as the dotted lines. Our 16 \name
implementation can forward at 100~Gbps
for every packet size other than 64 bytes, where it achieves 88\% of maximum
rate (or 125~MPPS). For 200~Gbps, the 16 \name implementation can forward 
at line rate for all packet sizes (note: below 128-byte, packets have reduced packet
generation performance). For 8 \names, we perform similarly for 100~Gbps,
but packets have to be at least 1024 byte packets to forward at the full 200~Gbps line rate. 

The performance drop for 64 and 65-byte packets is primarily 
due to the RISC-V software latency in the \names. For instance, the minimum time for our packet forwarder to read a
descriptor and send it back is 16 cycles. Therefore, each \name forwards a
packet every 16 cycles, and with 16 {\name}s we can hit at most 250~MPPS (equal
to the clock rate). Similarly, with 8 \name design we reach a maximum packet rate of 125~MPPS. 
For the single port forwarding, our distribution subsystem is limited to 125~MPPS per incoming
port limitation, which can be improved in the future.

\subsection{Forwarding Latency}
\label{sec:eval_latency}

Next, we measured \fname's forwarding latency. Namely, the round trip time (RTT) from
the traffic generator, through the FPGA under test, and back to the generator.
To measure the latency, the packets are time-stamped just before leaving the packet generator, and the
time is recorded upon arrival of them after the loopback. Figure~\ref{fig:latency_data} shows the measured
latency for different packet sizes, both under low-load and maximum-load
scenarios. The primary source of the measured delays is serialization. When a packet arrives
at the FPGA and when it leaves the FPGA, MAC FIFOs add serialization latency at
the line rate of each interface (i.e., 100~Gbps). \fname introduces additional serialization latency at 32~Gbps
due to the fact that packet is fully loaded into each \name's shared memory before the
RISC-V core is notified. The packets also have to be fully serialized on the way
out of the \name
after the descriptor is released.  The dotted line in
Figure~\ref{fig:latency_data} shows the computed serialization delay according
to the theoretical minimum latency given serialization
(Equation~\ref{eq:latency}). The 0.765$\mu$sec in this equation corresponds to the minimum latency of
packet forwarding through \fname, measured for the smallest packet size.

\vskip -1em
\begin{equation}
Est.\;latency\:(\mu s) = (size * 8 * (\tfrac{2}{100}+\tfrac{2}{32})/1000) + 0.765
\label{eq:latency}
\end{equation}
\vskip 0.5em

Since \lb tracks the available slots in the \names, any packet past the \lb can
be absorbed by \names. Therefore, high load introduces only marginal additional
latency, which likely comes from packets
not being uniformly distributed between the two outgoing interfaces. 
The only exception is for 64-byte packets, where our
packet generator can supply at a higher rate than what our packet forwarder can
support, causing the receive FIFO to become full in steady state and add
32.8$\mu$sec of additional latency.

\subsection{Inter-\name Messaging Performance}

Finally, we measured the throughput of the inter-\name loopback
messaging system. Our implementation only uses a single 100-Gbps
loopback port, since sending packets among cores for every incoming packet 
is not the intended design. 
To test the performance of this loopback port, we
implement a two-step forwarding system: we assigned half of the \names to
be recipients of the incoming traffic, and then each of these \names
forwards packets to its corresponding \name in the other half, and
finally that core returns the packet to the link. We achieve 60\% and 61\% maximum
throughput for the smallest packet sizes 64~Bytes and 65~Bytes respectively. Mainly
\fname is bottlenecked by the destination \name header having to be attached to each
packet. For packet sizes larger than 128 bytes, the system can keep up 
with the full 100-Gbps line rate.

We also performed two tests to measure the latency of broadcast messages: one 
with a fixed-rate of sparse messages, and one where each \name is trying to send as many messages
as fast as it can. We time-stamp each message by writing the time-stamp value in
the broadcast region, and upon arrival compare the current time against the transmit
time. In the normal scenario of sparse messages, we observed a latency between 72 to 92~ns. 
When trying to send as many messages as possible---which is not the intended use for this communication channel---we observe
1,596--1,680~ns of latency for the design with 16 {\name}s. This
latency mostly comes from the 18~FIFO slots in each \name---16 from actual FIFO
and 2 from PR registers---which can be sent out every 16 cycles due to
round-robin arbitration among cores (or every 8 cycles for the design with 8
{\name}s). This accounts for 1,152~ns of this latency:
$ 16\times18$ cycles, each at 4~ns. A write to the broadcast memory region will be blocked until there is
room in the FIFO. The rest of the latency is due to FIFOs and registers in the
packet distribution subsystem and the RISC-V software having slight variations.

\section{Case Studies}
\label{sec:casestudy}

In this section we demonstrate how \fname can reduce development effort and
time to build FPGA-accelerated middleboxes that can run at 200~Gbps. We
show this with two case studies.

\paragraph{Case study 1: Porting Pigasus to achieve 200G} For our first case
study, we ported Zaho et~al's Pigasus IDS~\cite{pigasus} to \fname. Pigasus
is the first open source FPGA hardware design to achieve 100~Gbps throughput. 
We show that we can use the \name abstraction to scale up the performance to twice the
rate they achieved in their paper. The main questions we answer in
this case study are as follows: How easily can we port the core Pigasus hardware
accelerators---string and port matching---to \fname's {\name}s?
Can we use \fname 
to achieve incremental hardware design where we start with a base accelerator and incrementally
add improvements and observe their performance gain?
Finally, how much the \fname framework can improve Pigasus's performance?

\paragraph{Case study 2: Building a Blacklisting Firewall.} For our second case
study, we evaluate how much easier \fname makes developing a new
middlebox from scratch. Namely, we show it is feasible to implement a simple 200~Gbps firewall 
based on a single hardware accelerator that blocks packets that match an IP blacklist.

We were able to develop both of these case studies in less than a month in
total by only one developer. We show \fname
improved the line-rate of the Pigasus IDS from 100~Gbps to 200~Gbps for
800~Bytes packets\footnote{The average packet size for internet traces is over 
800~Bytes~\cite{pigasus}.}, and achieves 200~Gbps for packets as small as 256~Bytes for the firewall implementation.

\subsection{Pigasus IDS/IPS}
\label{sec:pigasus}

\subsubsection {Why is an IDS Hard to Develop in an FPGA?}

IDSes identify suspicious behavior by monitoring network traffic and comparing
it to a set of known fingerprints, stored in a constantly evolving ruleset.
Many operators run all incoming traffic through an IDS, however they often have
to divide traffic across clusters of servers to handle the
computationally-intensive pattern matching for line-rate
traffic~\cite{ids-loadbalance}. This computationally expensive operation is easy
to parallelize; therefore, FPGAs are often considered for
accelerating IDSes~\cite{ids-100g,bro-fpga,snort-fpga}. The Pigasus team built
the first open-sourced FPGA-first IDS accelerator to provide 100~Gbps acceleration for the
Snort IDS running on a single server~\cite{pigasus}.

However, FPGA developers such as the Pigasus team had to build from scratch a significant
fraction of their IDS hardware design to hit line-rate on an FPGA.
The developers had to build
their own packet processing pipeline from scratch, including building hardware
accelerators for parts of the processing that could be done in software, such as
packet parsing.

\subsubsection{Porting Challenges and New Features}
\label{sec:pig_features}

The details of the steps required to port Pigasus to \fname are provided
in Appendix~\ref{app:pig_port}. 
Unfortunately, in the first pass of building Pigasus with \fname, the scaled-up
200~Gbps design did not fit in our FPGA. After reaching out to
the Pigasus team, they mentioned that lack of memory resources in the FPGA was a bottleneck for going to 200~Gbps,
even when they used a large Intel Stratix 10 MX FPGA. However, upon a close
look at our resource requirement report, we noticed that no large URAMs were used for
the large lookup tables in string or port matcher accelerators.
This is because URAMs cannot be initialized when an FPGA bitstream
is loaded, as they are targeted for FIFOs. One method would be to initialize
them during runtime, but before \fname that would have required more development effort to enable initialization from the server hosting the FPGA.

Using the \name memory subsystem in \fname, we were able to fill these tables at
runtime. We simply added a write port to the four large lookup memories in the
accelerators. Still full Pigasus string matching engine did not fit
in a single \name, as they previously used the full FPGA for it. Fortunately,
the number of string matching engines was parameterized, and by just changing it to 16
from 32, in addition to benefiting from the footprint reduction using URAMs, it fit
within an \name. The next highest utilization resource was the DSP blocks used for
computing the hash for addressing the table, but it was still within the
available capacity. We used the layout with 8 \names, as it provides more
resources per \name. A layout with 4 \names would have more resources
per \name, but the overhead of software running on RISC-V cores would become a
bottleneck.

\fname also enabled overcoming a key limitation of the original Pigasus design: there is no way to reconfigure
the pattern matcher's ruleset during runtime. The only method to update the
ruleset is to reload a new FPGA image. Not only we can benefit from the PR
support in \name, but also we can use the packet distribution subsystem to
modify the large lookup tables during runtime.

\begin{table}
\caption{Average resource utilization for \names with Pigasus, and the accompanying Hash-based \lb}
\label{table:pigasus_utilization}
\vskip -1em
\centering
\resizebox{\linewidth}{!}{%
\setlength{\tabcolsep}{4pt} 
\begin{tabular}{|c|c|c|c|c|c|}
  \hline
  Component       & LUTs            & Registers     & BRAM         & URAM & DSP  \\
  \hline
  RISCV core      &  2048 (3.2\%)   &  1051 (0.8\%)  & 0            & 0 & 0         \\
  Mem. subsystem  &  3503 (5.5\%)   &  906  (0.7\%)  & 16 (14.0\%)  & 32 (50.0\%) & 0 \\
  Accel. manager  &   803 (1.2\%)   & 2717  (2.1\%)  & 0            & 0           & 0 \\
  Pigasus         & 36012 (56.1\%)  & 49364 (38.3\%) & 56 (49.1\%)  & 22 (34.4\%) & 80 (20.8\%) \\
  Total           & 42364 (66.0\%)  & 54037 (41.9\%) & 72 (63.2\%)  & 54 (84.4\%) & 80 (20.8\%) \\
  \hline
  \name           &  64161          &  128880        & 114         & 64   & 384      \\
  \hline
  \lb              &   10467 (0.9\%) &   24872 (1.0\%) &  26 (1.2\%) & 0 & 0 \\
  Remaining        &  103549 (8.8\%) &  205528 (8.7\%) &  154 (7.1\%) & 96(10.0\%) & 648 (9.5\%)  \\
  \hline
\end{tabular}
}
\vskip 0.2em
\end{table}

To further demonstrate the benefits of the bridging between software and hardware that \fname provides, we also
implemented the Pigasus's TCP flow reordering in software on the RISC-V
core, instead of porting their hardware accelerator. We built a hash-bashed \lb for \fname to
always send packets of the same flow to the same \name. 
The \lb also pads the 4-byte hash result to the beginning of each packet, so
that the software can reuse the flow hash without recomputation, and
also know the exact hash that the \lb has used.
In each \name, we used the scratch pad memory to keep 0.5~MB of flow state:
storing the time and sequence number of the last seen packet
from each flow, the flow hash, and the last 7
bytes of the packet required to be checked for the next packet. 
We can fit 32K entries of 16~Bytes within the 0.5~MB \name memory, 
covering 15~bits of the hash as the index to this table. Since the \lb
has split the flows based on 3~bits of the same hash, in total we have covered 18~bits
of the hash (out of 32~bits). Furthermore, older flows 
quickly time out in our software implementation, making a 
hash collision very unlikely. 
If we encounter reordering, we use up to half of our packet slots
(e.g., 16) to buffer the out-of-order packets until the re-ordered packets
arrive. 
In the rare case of collision, or running out of reordering buffers, we forward
the corresponding packets to the host. 
Table~\ref{table:pigasus_utilization} shows the average resource utilization break down
inside each Pigasus \name unit, as well as the hash-based \lb used
for this mechanism.

The software reordering will be less efficient than the hardware-based solution
proposed in Pigasus, but utilizing software next to hardware can be used for
prototyping or testing new ideas. To better understand the overhead of software
reordering, we evaluate \fname assuming such hardware reordering accelerator
exists within our round robin \lb, similar to the inline hash computation accelerator
within our hash-based \lb. This is reasonable as their reassembler accelerator
keeps the state per flow, and attaches the required state to each packet, so no
state needs to be kept within \names and any \name can process any packet. Since
the addition of this accelerator would not impact the performance of software
within \names, porting it was not necessary. However, compared to the numbers
reported in the Pigasus paper, plenty of available resources still remain in our
round robin \lb (as shown in Table~\ref{table:utilization_8}), should someone
choose to add the reorder accelerator in the future.

\subsubsection{Performance Evaluation}

\begin{figure}
\centering
  \subfloat[\label{fig:pigasus_rate}]{%
    \includegraphics[width=0.75\linewidth,trim={0 1em 0 0},clip]{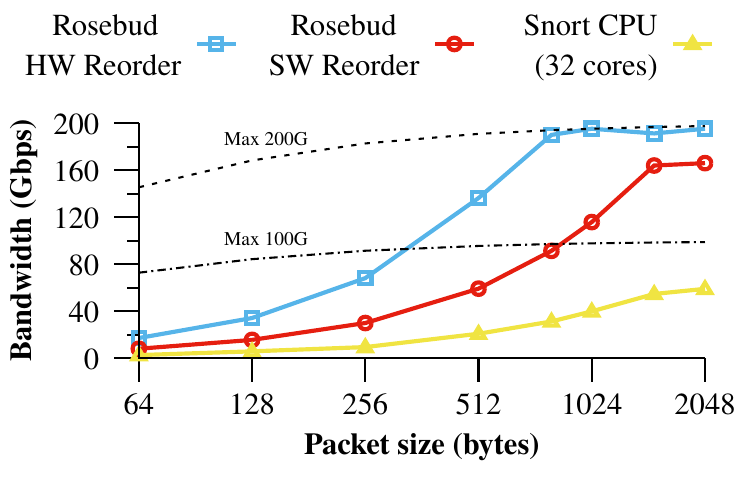}}
  \\
  \subfloat[\label{fig:pigasus_pps}]{%
    \includegraphics[width=0.75\linewidth,trim={0 1em 0 0},clip]{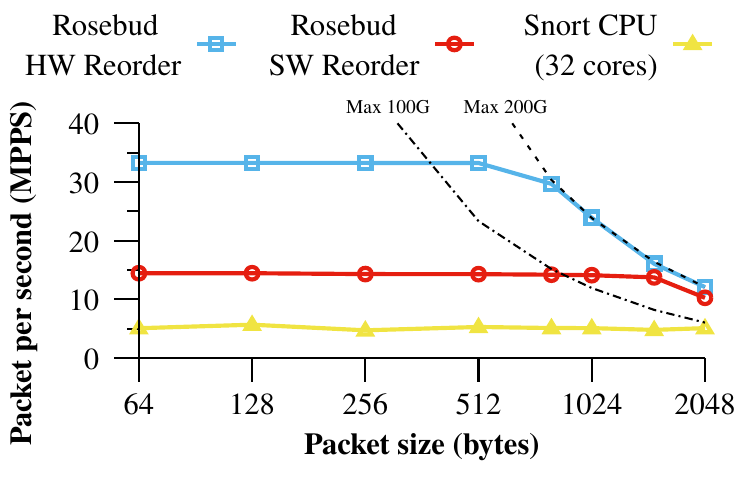}}
\caption{
\label{fig:pigasus_res}
IPS performance in terms of (a) bandwidth and (b)
packet rate, using Pigasus in \fname, compared with Snort.}
\end{figure}

To evaluate the performance of Pigasus on \fname, we first made a packet trace based on the ruleset used for
the generation of the Pigasus accelerator. Then, we used \texttt{tcpreplay} to transmit packets at low rate to
verify that the expected number of packets are selected by the IPS accelerator and sent
to the host. Next, using tcpreplay we played back the attack pattern, and used
another FPGA to fill the rest of the pipe with background traffic to hit
200~Gbps. For this case study we ran the attack as 1\% of the traffic (2~Gbps)
with 0.3\% reordering among the TCP flows in both the attack and safe traffic.
This attack rate is considered a medium to high rate, and the reordering rate is
the typical reordering happening for middlebox traffic~\cite{pigasus}.

By using this traffic generation setup, we compared performance of three
systems: (1) Pigasus with hardware accelerated reorder engine, (2) Pigasus with
software reorder engine running on RISC-V cores, and (3) Snort running on a
machine with Intel Xeon 6130 with 32 physical cores---or 64 hyper-threaded
cores. For the Snort performance measurements, we setup the rules so Snort is
performing only the exact same fast-pattern matching 
performed by the Pigasus accelerators. Also we enabled Snort's
Hyperscan~\cite{hyperscan} to improve fast pattern matching speed by benefiting
from Intel's AVX-512 vector instructions within Intel CPU, as well as the Linux
Kernel's AFPACKET~\footnote{\url{https://doc.dpdk.org/guides/nics/af\_packet.html}}
to minimize the overhead for moving packets from the kernel to userspace.
Enabling both of these resulting in more than 2 times improvement in 
performance.

Figure~\ref{fig:pigasus_rate} shows the results of this experiment. \fname with 
both hardware acceleration for both reordering and pattern matching achieves
the highest performance: almost 200~Gbps for packet sizes larger than 800~Bytes.
The use of software running on the slow RISC-V cores to do reordering
does indeed lower performance, but it still can achieve almost 100~Gbps at
800~Bytes, and 166~Gbps for 2048~Byte packets. The non-perfect load
balancing among the \names, due to non-uniformity of the flow hash results, also plays
a smaller part in the performance degradation. Snort performs worse than both
\fname implementations.

We also evaluate the performance differences from the perspective of packet rate 
(Figure~\ref{fig:pigasus_pps}). The packet rate is limited by the
software running on RISC-V cores in \fname 
to 60 and 138~MPPS for the HW-based and SW-based 
reodering, respectively. That is the limiting factor until 512~Byte
packets for the HW-based reordering, and 1024~Bytes packets for the SW-based.
After this thresholds, another limitation is added which is the maximum packet rate for
a specific packet size at 200~Gbps, which is depicted by the dotted lines.
For the HW-based reordering, from 800~Bytes packets the max packet rate is the
dominant limiting factor and sets the performance, while for the SW-based
results the software overhead still plays a part, even for 2048~Bytes packets.

For the Snort performance on a multi-core CPU, we see that for different
packet sizes, the packet rate is limited between 4.7 and 5.6 MPPS. Although
Snort is running on Xeon cores which have an order of magnitude higher clock rates
than the FPGA,
and Hyperscan hardware acceleration does speed up pattern matching 
on the multicore CPU, the
achieved packet rate is much lower than \fname on an FPGA. This may be because
Hyperscan's acceleration is done in a sequential
manner, even the AVX-512 instructions are not as fully pipelined, compared to the pattern matching in the Pigasus FPGA implementation.
To make sure the Linux network stack is not introducing a bottleneck for Snort, we did an additional
experiment where we ran Snort on only safe traffic with 2048~Bytes packets 
read directly from the RAM (Using ramdisk). The
performance increased from 60~Gbps to only 70~Gbps, proving that 
AFPACKET moving packets between the interface and Snort is not the primary bottleneck for Snort's performance.
These results also clearly motivate why Pigasus can achieve >100~Gbps on a
single host: the FPGA filters non-attack traffic coming in at line-rate, and
the CPU only deals with attack traffic at a fraction of the line-rate.

\subsubsection{Performance Analysis of Pigasus on \fname}

\fname achieved twice as the performance of original Pigasus due to increased amount of
parallelization. We used 16 string matching engines inside
each \name, as opposed to $32$ in the original Pigasus's design. But we have 8
\names within our FPGA, that makes our system 4~times more parallel.
Table~\ref{table:pigasus_utilization} shows the resource utilization for this
integration. As a point of comparison and benefits from using URAM memories,
These 8 \names are placed within 2/3rd of our FPGA which in total has a similar
capacity to their FPGA.

Considering that there were 8 \names and within each \name the accelerators could
potentially process 16~Bytes per cycle, the maximum rate for this computation at
clock frequency of 250~MHz would be 256~Gbps. This is more than the target
200~Gbps performance, and hence not a bottleneck. However, as observed in
Figure~\ref{fig:pigasus_rate} the 200~Gbps is not achievable in many packet
sizes. This results are also lower than the throughput of the packet distribution system,
as shown in Section~\ref{sec:eval}, indicating that packet distribution is not the bottleneck either. 
Indeed the software
running on the slow RISC-V cores is setting the limit for small packet sizes.

\begin{figure}
  \centering
  \includegraphics[width=0.7\linewidth]{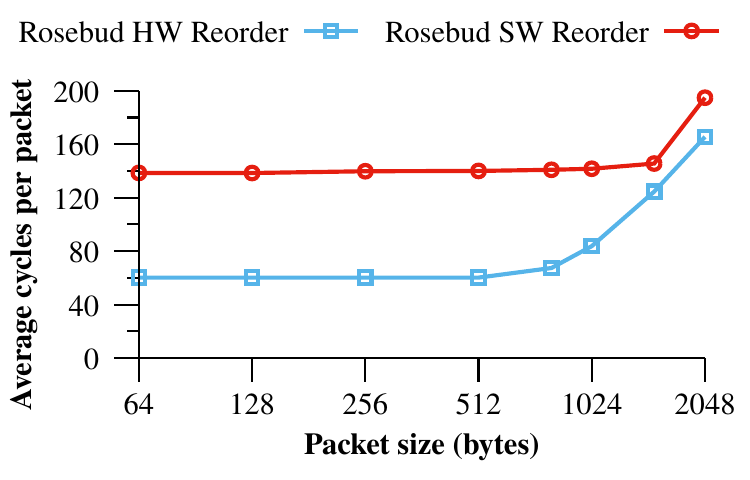}
  \caption{
  \label{fig:pigasus_cycles}
  Average cycles spent per packet
  }
\end{figure}

Based on the measured packet rate,
we
can compute the number of cycles spent per packet, depicted in
Figure~\ref{fig:pigasus_cycles}. When no reordering is performed in the software
(HW reorder line), the cycles spent per packet is 60.2 for the small packets. When we
checked the simulation results for the same packet size and the same C-code, we
observed that it takes 61 cycles for safe TCP packets, 59 cycles for safe UDP
packets, and 82 cycles for attack traffic (the full code can be found in
Appendix~\ref{app:pigasus}). Therefore, on average, for 1\% attack
rate and a small portion of total packets being UDP, the measured results
matches our simulation. 

For hardware reordering, the software only runs
the parsing and accelerator management, which could be fully parallelized with
the Pigasus accelerator, therefore even for packet size 1024~Bytes which requires
61 cycles for Pigasus to process the payload, the simulation results were still
61 cycles, and only at 2048~Bytes did the Pigasus runtime become the
bottleneck. That being said, for packet sizes 800 and above the maximum packet
rate at 200~Gbps is bottleneck, not the software on RISC-V core.

For software reordering, the software needs to also handle the flow state table, 
resulting in more cycles spent per packet. The average cycles per packet starts
from 138.4 cycles at 64~Bytes, and slightly rises until 1500~Bytes, where the
line rate becomes the bottleneck and measurements do not reflect the software
overhead anymore. This increase is due to less overlapping opportunity for
the management software and the hardware accelerator, as starting off the Pigasus
accelerator is dependent on the flow state table read results in software. 
For the software reordering case, the simulation results were not as
consistent as before, but they were around the measured average value. To get a
proper value it required potentially simulation of thousands of packets, which is
not practical.

Additionally, during these experiments we learnt that careful software design and compiler
efficiency plays a critical role to achieve high performance in \fname. 
For instance, for the implementation with hardware reordering, we obtained a
30\% improvement in packet rate by adjusting the order of members in a struct,
and also applying a bug fix from the latest RISC-V GCC which was not available
on the Arch Linux repository version at the time.

\subsection{Firewall}

We next investigated how much developer effort is required to build a simple firewall middlebox from scratch using \fname.
A firewall checks every single packet, and drops the packets whose IP matches a
blacklist, otherwise they are forwarded to the other
Ethernet interface. 

To implement a firewall, we built a simple IP prefix lookup hardware accelerator from the list
of 1050 blacklist IPs in the ``emerging
threats\footnote{\url{https://rules.emergingthreats.net/fwrules/emerging-PF-DROP.rules}}'' firewall rules.
We wrote a basic python script to parse the rules and generate a Verilog code
for the accelerator. This accelerator first checks for the
first 9 bits of the IP prefix, if they match, then it checks for the remaining 15
bits in the next cycle, and if there was a match it raises a flag in a
register. This lookup can be performed in 
only two clock cycles.

Then we assigned a register address that the RISC-V core could use to load the
IP into the accelerator using MMIO, and another register to read the flag.
Listing~\ref{fw_snippet} shows a small code snippet from the RISC-V code to show how this works 
(the full code can be found in the Appendix~\ref{app:firewall}).
We load the IP address from the Ethernet packet using the DMA descriptor pointer given by
the \name interconnect. Then we load the address into the IP matching accelerator
(\textsc{ACC\_SRC\_IP}) and check the flag to see the results
(\textsc{ACC\_FW\_MATCH}). In case of a match, we drop the packet by
setting the descriptor length to 0; otherwise, we forward it by
swapping the port value between 0 and 1; asking the packet distribution subsystem to send the packet to the other 100~Gbps port.

\vspace {4pt}
\begin{lstlisting}[style=CStyle,caption={RISC-V code snippet for the firewall case study}, label=fw_snippet]
  unsigned int src_ip =
      *((unsigned int*) (desc->data + 14 + 12));
  ACC_SRC_IP = src_ip;
  if (ACC_FW_MATCH) desc->len   = 0;
  else              desc->port ^= 1;
  pkt_send(desc);
\end{lstlisting}

We were able to hit 200~Gbps for packets 256~Bytes and above, while injecting
attack traffic within the background traffic. Table~\ref{table:firewall_utilization} shows the average resource utilization breakdown
inside each firewall \name unit. The resource consumption per firewall
engine is small and for more number of rules, several such engines can be used
in parallel and coordinated by the software.

\begin{table}
\caption{Average resource utilization for \names with firewall}
\label{table:firewall_utilization}
\vskip -1em
\centering
\resizebox{\linewidth}{!}{%
\setlength{\tabcolsep}{4pt} 
\begin{tabular}{|c|c|c|c|c|c|}
  \hline
  Component       & LUTs            & Registers     & BRAM         & URAM & DSP  \\
  \hline
  RISCV core      &  1976 (7.1\%)   & 1050 (1.9\%)  & 0            & 0 & 0         \\
  Mem. subsystem  &  2166 (7.8\%)   &  862 (1.5\%)  & 16 (44.4\%)  & 32 (100\%) & 0 \\
  Accel. manager  &  518  (1.9\%)   & 1944 (3.5\%)  & 0            & 0          & 0 \\
  Firewall IP checker & 835 (3.0\%) &   197 (0.4\%)  & 0    & 0    & 0          \\
  Total           &  5493 (19.7\%)  & 4053 (7.3\%) & 16 (44.4\%)  & 32 (100\%) & 0 \\
  \hline
  \name          &  27839           &  55920        & 36         & 32   & 168      \\
  \hline
\end{tabular}
}
\vskip -0.3em
\end{table}

\section{Related work}
\label{sec:related}

There has been a significant amount of prior work in developing FPGA-based NIC and also 
System-on-Chip (SoC) NIC development frameworks that are flexible and debuggable. However, no
prior framework has addressed the issues 
for FPGA middleboxes, while providing a software like development and debugging, and
the ability to change hardware acceleration---even during runtime.

\subsection{FPGA Frameworks for NICs and Switches}
This work builds on a large body of prior work on improving the software and
hardware development process for FPGA-based NICs. However, they were not targeted at the
specific developer needs of FPGA middleboxes.
The HxDP~\cite{hxdp} framework, demonstrated a
significant number of the development features could be provided by one
framework.
However, HxDP focuses on software-based compilation development
environment for FPGAs and does not provide an interface to add new hardware
accelerators to their framework; therefore it also does not provide a
solution to parallelize custom accelerators.
The PANIC~\cite{panic2} NIC framework has a similar packet distribution
system to \fname; however, it does not provide software control over hardware
accelerators, nor the software-like debugging. Also it is mainly targeted for
ASIC implementations.

Other frameworks provide a well-defined shell on an FPGA to add custom accelerators~\cite{netfpga, netfpga-sume, risc-netfpga, catapult-accelnet,
catapult, clicknp, gorilla}.
However, none of these 
provided a software-based development
process.
There are also FPGA frameworks that are tailored for applications written in the P4
language~\cite{p4fpga,flowblaze,p4vbox}, but they are targeted for only switching
applications rather than middleboxes, and do not provide software-oriented debugging
capabilities like \fname. 
None of these prior frameworks demonstrated it was feasible to build a generic
middlebox FPGA abstraction, like \fname, that can provide these features simultaneously 
and also achieve line-rate performance.

\subsection{HW-Accelerated Network Applications}

There are numerous projects that implemented hardware accelerated network
middleboxes with their own custom FPGA hardware pipeline, including
an ML platform~\cite{catapult-deeplearning}, a key value store~\cite{kvdirect}, packet filtering~\cite{ffshark}, and several Intrusion Detection
Systems~\cite{snort-fpga,netfpga-osnt,bro-fpga,ids-100g,retina}. We believe that future
efforts such as these may be bolstered by this platform. With \fname, developers will
be able to focus on building their application-specific accelerators,
and not have to manually tune a pipeline to get high-performance
or manually build debugging hardware.

\section{Conclusion and Discussion}
\label{sec:concl}

\fname is a flexible and debuggable FPGA middlebox development
framework. It provides a packet-processing abstraction consisting of
a RISC-V core augmented with hardware accelerators, unified by
custom tailored shared memory and packet distribution subsystems.
We demonstrate that \fname can achieve
200~Gbps. We also demonstrate that \fname has a marginal effect on
latency (especially when compared with PCIe and OS latencies). We plan to port
\fname to several FPGA boards, from both Xilinx and Intel, to make it possible to
use the same {\name} abstraction among them. 
\fname can also be used for sharing FPGAs in the cloud services, such as
Amazon AWS-F1~\cite{AWS-F1}, where the cloud provider controls the \lb and
users can load their logic into the \names.
Finally, although FPGA-based middleboxes benefit most from the
flexibility offered by \fname, we believe scope of \fname can be potentially
wider. SoC-based SmartNIC designs can benefit from the memory and messaging
subsystems to scale to higher link speeds. 
Fully custom ASIC designs can use \fname for their
incremental builds where only the accelerators are updated between revisions. 

\paragraph{Discussion: Rosebud on Hybrid FPGA platforms} 
A potential platform for \fname is SoC-like FPGAs with hardened CPUs, such
as Xilinx Zynq UltraScale+ MPSoC~\cite{zynq}. However, they have limited
parallelism with a limited number of cores, small memory per core, and low incoming
bandwidth ($<20$~Gbps per core). Most importantly though, they have high
communication latency to the logic fabric
($>100$~ns~\cite{zynq-latency1,zynq-latency2}) in the
FPGA and use a generic shared bus that introduces contention, both of which
critically limit their ability to orchestrate parallel processing in the
accelerators. 
The \name-based framework can be ``hardened'' in
an SoC architecture in future FPGAs. Xilinx already has Network-on-Chip
IP in their Versal family of FPGAs~\cite{versal-premium} that can save the resources used for
packet distribution. AI cores available in this family are also similar to a software based
processor, but they are not a full C-based core and also they are not spread
across the programmable logic. 
Hardening the RISC-V cores will result in significantly faster
packet processing performance, compared to the ones made on top of  programmable logic. 
\fname already fully supports clock crossing between the core
and accelerators domains.
Furthermore, for applications that require large memories, the
Versal family also includes FPGAs with integrated High Bandwidth Memory (HBM) up
to 32~GB. If even larger memories are required, \fname can be mapped to Intel
Xeon-FPGA hybrid chips~\cite{xeon-fpga} to have faster access to host memory. 

\begin{acks}
We thank Manya Ghobadi for supporting this project while Moein was doing his
postdoc at MIT. Also, we thank the anonymous reivewers for their detailed
feedback. The MIT-affiliated author was supported by ARPA-E ENLITENED PINE
DE-AR0000843, DARPA FastNICs 4202290027, NSF CNS-2008624, NSF SHF-2107244, NSF
ASCENT-2023468, NSF CAREER-2144766, NSF PPoSS-2217099, NSF CNS-2211382, Meta
faculty award, Google faculty award, Juniper Networks Sponsored University
Research Initiative (SURI) award, and Sloan fellowship FG-2022-18504.
\end{acks}

\label{lastpage}

\clearpage
\appendix 

\section{Steps to port Pigasus accelerator} 
\label{app:pig_port}

In this appendix, we describe the steps to port a design to \fname, following
our process for the Pigasus case study. We chose to port two of the main
hardware accelerators from Pigasus: the pattern matcher, and the port matcher.
As discussed in Section~\ref{sec:pig_features}, 
we skipped porting their reassembly engine and modeled it by our round robin \lb
itself, while we replicated the functionality with a hash-based \lb and use of
software running on the RISC-V.

\subsection{Developing the Accelerator} 
First, we simply copied the files in the string pattern matcher and port matcher
directories from the Pigasus code base. We only had to swap a few IP 
modules---such as FIFOs---which were generated with Intel Quartus and were not
compatible with Xilinx Vivado. We used our own libraries written in Verilog instead. Then we
set the parameter for their rule packer module to output in chunks of 32
bits (rather than 128 bits) to match our RISC-V word size. 

\subsection{Connecting the Accelerator to \name}
Next, we had to write a basic wrapper to connect the Pigasus accelerators to
the system within an \name. The role of this wrapper is mainly to connect the wires and
set the desired addressing for configuration registers.  Their accelerators use
a streaming input and could be connected directly within an \name, as we support both the
streaming data and random access memory---also known as native memory---interfaces. 
For the address assignment logic, it is a basic
case statement in Verilog, and we provide examples for both blocking or non-blocking read
and writes from the software running on RISC-V to the registers.  
Furthermore, to reduce the overhead of software control, we add basic hardware
queues (FIFOs) per accelerator in this wrapper. Such queues make the orchestration of
accelerators similar to an asynchronous scheduling software that manages local
resources, where the software feeds and checks the accelerator resources.

\subsection{Writing the Accompanying C Code}
To have a complete \name, we also need the software for the RISC-V processors to support
the accelerators similar to a firmware.  We use GNU RISC-V GCC to compile the
code alongside the provided libraries to receive and send packets, and
communicate with different components of \fname and the host. This
code parses the packet headers, manages the Pigasus accelerator by
feeding the start pointer and length of the payload, and finally
appends the matched rule IDs to the matched packets before sending them to the host, or
sends the safe traffic out on a physical port.

\subsection{Simulating the C Code with Accelerators}
At this stage, we can proceed to simulation to verify the interaction between the
C code and the accelerators.  We provide a Python-based simulation framework,
using CocoTB~\cite{cocotb} simulator to connect Python to an RTL simulator
for the Verilog code, Use of Python considerably simplifies
writing test benches. For example, we could easily use Python libraries such as
Scapy~\cite{scapy} to craft packets for testing, and also use
idstools~\cite{idstools} to parse the rules and make the attack packets
accordingly. To run the simulation, we provide Python functions to load the
memories of the \names from the outputs generated by the
GCC, as well as accompanying function to send and receive packets to the rest of
the system. We provide both options of single \name or full \fname
simulation, the latter being more complete but also more time-consuming. At this stage,
single \name simulation proved to be more useful, as we were elaborating the
interaction within an \name, and also the packet distribution framework
practically isolates the functionality of each \name. The full \fname simulation
can become handy to elaborate the design of a new \lb, or communication between the
\names if necessary.

\subsection{Generating the Bitstream for FPGA}
After verifying the functionality of the Pigasus accelerator within \fname, we
proceeded to implementation with the FPGA development toolchain.  We only needed
to add the Verilog file names of the Pigasus accelerators, and the corresponding
accelerator wrapper we developed, to our script and let the tool chain output the
bit file for the target FPGA. Regions for \names are already laid
out, and the rest of the logic is already placed and routed, so we only have to
build the bitstreams for the partially reconfigurable regions of each \name.
After the first implementation run, we ran into the problem of insufficient
resources per \name. This issue was addressed through the additional capabilities 
provided in \fname framework, discussed in Section~\ref{sec:pig_features}.

\subsection{Load the FPGA and \name Memories}
Next, we loaded the bitstreams on the target FPGA, using our Bash scripts. We
developed a host-side C library to communicate with the FPGA.  This library
integrates the Xilinx's PR-loading tool, \texttt{MCAP\_tool}, and the Corundum
100~Gbps FPGA-to-host NIC driver~\cite{corundum}. Therefore, we update an
\name using the PR-loading, and initialize the {\name}s memories and \lb
configurations through C code running on the host. We used this library to load
the instruction, data and accelerator memories of the \name, directly from the ELF
output file of GCC.

\subsection{Runtime Debugging}
For debugging, we use the host-side libraries where we have read and write
access to the memories within an \name, and we can send interrupts or read
status bits from each \name.  We also developed a separate 64-bit debug channel
between the host and each \name. In a deadlock scenario, host can send a poke
interrupt and communicate via an \name in both directions using this channel at a 
lower rate.

\subsection{Runtime Reconfiguration}
To do reconfiguration of an \name at runtime, we send a signal from the host to
the \lb to stop sending packets to that specific \name, and then send an
eviction interrupt to the RISC-V core to instruct it to finish processing the
current packets and save the desired state to the host. Next we use the host-side
libraries to write the new bit file to the PR block corresponding to the target
\name, followed by loading the \name's instruction and data memories. Now we
reset the core and let it boot up, and restore its state by reading from the host
memory if necessary.  When the core is ready, we signal the \lb through the host to 
resume sending packets to that \name again.

\onecolumn

\begin{multicols}{2}

\section {Code for Pigasus case study}
\label{app:pigasus}

The following code shows the C code running on the RISC-V cores for the Pigasus case
study. This code is for the hardware reorder test and does not keep the flow state.
The full code for the software reorder implementation can be found in the repository.

\end{multicols}

\vskip -1em

\begin{lstlisting} [style=CStyle]
  #include "core.h"
  #include "packet_headers.h"
  
  #define bswap_16(x) (((x & 0xff00)     >> 8)  | ((x & 0x00ff)     << 8))
  #define bswap_32(x) (((x & 0xff000000) >> 24) | ((x & 0x00ff0000) >>  8) \
                     | ((x & 0x0000ff00) <<  8) | ((x & 0x000000ff) << 24))
  
  #define mem_align(x) (((unsigned int)x+3) & 0xFFFFFFFC)

  // maximum number of slots (number of context objects)
  #define MAX_CTX_COUNT 32
  #define REORDER_LIMIT 8
  
  // Packet start offset to DWORD align Ethernet payload and provide space for header modifications
  #define PKT_OFFSET 10

  // PMEM in 8 blocks of 128 KB Accelerators are only connected to upper 2 blocks
  #define PKTS_START ((8-(MAX_CTX_COUNT/8))*128*1024)
  
  // Accel wrapper registers mapping
  #define ACC_PIG_CTRL     (*((volatile unsigned char      *)(IO_EXT_BASE + 0x00)))
  #define ACC_PIG_MATCH    (*((volatile unsigned char      *)(IO_EXT_BASE + 0x00)))
  #define ACC_PIG_STATE    (*((volatile unsigned long long *)(IO_EXT_BASE + 0x10)))
  #define ACC_PIG_STATE_L  (*((volatile unsigned int       *)(IO_EXT_BASE + 0x10)))
  #define ACC_PIG_STATE_H  (*((volatile unsigned int       *)(IO_EXT_BASE + 0x14)))
  #define ACC_PIG_PORTS    (*((volatile unsigned int       *)(IO_EXT_BASE + 0x0c)))
  #define ACC_PIG_SRC_PORT (*((volatile unsigned short     *)(IO_EXT_BASE + 0x0c)))
  #define ACC_PIG_DST_PORT (*((volatile unsigned short     *)(IO_EXT_BASE + 0x0e)))
  #define ACC_PIG_SLOT     (*((volatile unsigned char      *)(IO_EXT_BASE + 0x18)))
  #define ACC_PIG_RULE_ID  (*((volatile unsigned int       *)(IO_EXT_BASE + 0x1c)))
  
  #define HASH_LOOKUP      (*((volatile unsigned short     *)(IO_EXT_BASE + 0x60)))
  #define HASH_BLOCK_32B   (*((volatile unsigned char      *)(IO_EXT_BASE + 0x64)))
  
  #define ACC_DMA_LEN      (*((volatile unsigned int       *)(IO_EXT_BASE + 0x04)))
  #define ACC_DMA_ADDR     (*((volatile unsigned int       *)(IO_EXT_BASE + 0x08)))
  #define ACC_DMA_STAT     (*((volatile unsigned int       *)(IO_EXT_BASE + 0x78)))
  #define ACC_DMA_BUSY     (*((volatile unsigned char      *)(IO_EXT_BASE + 0x78)))
  #define ACC_DMA_DONE     (*((volatile unsigned char      *)(IO_EXT_BASE + 0x79)))
  #define ACC_DMA_DONE_ERR (*((volatile unsigned char      *)(IO_EXT_BASE + 0x7a)))
  
  // Slot contexts
  struct slot_context {
    struct Desc desc;
    int index;
  
    // Pointers
    unsigned char  *eop;
    unsigned char  *header;
  
    struct eth_header *eth_hdr;
    union {
      struct ipv4_header *ipv4_hdr;
    } l3_header;
    union {
      struct tcp_header *tcp_hdr;
      struct udp_header *udp_hdr;
    } l4_header;
  };
  
  struct slot_context context[MAX_CTX_COUNT+1];
  unsigned int pkt_num, slot_count, header_slot_base;
  
  const unsigned int slot_size        = 16*1024;
  const unsigned int header_slot_size = 128;
  
  static inline void slot_rx_packet(struct slot_context *slot)
  {
    unsigned int   payload_offset;
    unsigned int   packet_length  = slot->desc.len;
  
    // check eth type
    if (slot->eth_hdr->type == bswap_16(0x0800))
    {
      // IPv4 packet, check protocol
      if (slot->l3_header.ipv4_hdr->protocol == 0x06) // TCP
      {
        payload_offset = ETH_HEADER_SIZE + IPV4_HEADER_SIZE + TCP_HEADER_SIZE;
  
        ACC_DMA_ADDR  = (unsigned int)(slot->desc.data)+payload_offset;
        ACC_DMA_LEN   = packet_length - payload_offset;
        ACC_PIG_PORTS = * (unsigned int *) slot->l4_header.tcp_hdr; // both ports
        ACC_PIG_STATE_H = 0x01FFFFFF;
        // ACC_PIG_STATE_L = 0xFFFFFFFF; // Redundant
        ACC_PIG_SLOT  = slot->index;
        ACC_PIG_CTRL  = 1;
        return;
      }
      else // UDP
      {
        payload_offset = ETH_HEADER_SIZE + IPV4_HEADER_SIZE + UDP_HEADER_SIZE;
  
        ACC_DMA_ADDR  = (unsigned int)(slot->desc.data)+payload_offset;
        ACC_DMA_LEN   = packet_length - payload_offset;
        ACC_PIG_PORTS = * (unsigned int *) slot->l4_header.udp_hdr; // both ports
        ACC_PIG_STATE_H = 0;
        ACC_PIG_SLOT  = slot->index;
        ACC_PIG_CTRL  = 1;
        return;
      }
    }
    slot->desc.len = 0;
    pkt_send(&slot->desc);
  }
  
  static inline void slot_match(struct slot_context *slot){
    unsigned int rule_id;
  
    while (1){
      rule_id = ACC_PIG_RULE_ID;
      asm volatile("" ::: "memory");
  
      if (rule_id!=0){
        ACC_PIG_CTRL = 2; // release the match
        asm volatile("" ::: "memory");
        // Add rule IDs to the end of the packet
        slot->eop = (unsigned char *) mem_align(slot->desc.data + slot->desc.len);
        * (unsigned int *) slot->eop = rule_id;
        slot->desc.len = (unsigned int) slot->eop - (unsigned int) slot->desc.data + 4;
        slot->desc.port = 2;
      } else { // EoP
  
        ACC_PIG_CTRL    = 2; // release the EoP
        asm volatile("" ::: "memory");
        pkt_send(&slot->desc);
        return; // Go back to main loop when done with a packet
      }
  
      if (ACC_PIG_MATCH) // continue draining FIFO
        slot = &context[ACC_PIG_SLOT];
      else
        break;
    }
  }
  
  int main(void)
  {
    struct slot_context *slot;
    unsigned int reorder_mask, reorder_left_mask, init_left_mask;
  
    DEBUG_OUT_L = 0;
    DEBUG_OUT_H = 0;
  
    // set slot configuration parameters
    slot_count       = 32;
    header_slot_base = DMEM_BASE + (DMEM_SIZE >> 1);
  
    if (slot_count > MAX_SLOT_COUNT)
      slot_count = MAX_SLOT_COUNT;
  
    if (slot_count > MAX_CTX_COUNT)
      slot_count = MAX_CTX_COUNT;
  
    // Initializing LB and RPU interconnet
    init_hdr_slots(slot_count, header_slot_base, header_slot_size);
    init_slots(slot_count, PKTS_START+PKT_OFFSET, slot_size);
    set_masks(0x30); // Enable only Evict + Poke
  
    // init slot context structures
    for (int i = 1; i <= slot_count; i++)
    {
      context[i].index     = i;
      context[i].desc.tag  = i;
      context[i].desc.data = (unsigned char *)(PMEM_BASE + PKTS_START + PKT_OFFSET + (i-1)*slot_size);
      context[i].header    = (unsigned char *)(header_slot_base + PKT_OFFSET + (i-1)*header_slot_size);
      context[i].eth_hdr   = (struct eth_header*)(context[i].header);
  
      context[i].l3_header.ipv4_hdr = (struct ipv4_header*)(context[i].header + ETH_HEADER_SIZE);
      context[i].l4_header.tcp_hdr  = (struct tcp_header*) (context[i].header + ETH_HEADER_SIZE + IPV4_HEADER_SIZE);
    }
  
    ACC_PIG_STATE_L = 0xFFFFFFFF;
    // pkt_num = 0;
  
    while (1)
    {
      // check for new packets
      if (in_pkt_ready())
      {
        // compute index
        slot = &context[RECV_DESC.tag];
  
        // copy descriptor into context, we already know the data pointer
        slot->desc.desc_low = RECV_DESC.desc_low;
        asm volatile("" ::: "memory");
        RECV_DESC_RELEASE = 1;
  
        // handle packet
        slot_rx_packet(slot);
      }
  
      if (ACC_PIG_MATCH) {
        slot_match(&context[ACC_PIG_SLOT]);
      }
    }
  
    return 1;
  }
\end{lstlisting}

\vskip -1em

\begin{multicols}{2}

\section {Code for firewall case study}
\label{app:firewall}

The following code shows the C code running on the RISC-V cores for the firewall case
study.

\end{multicols}

\vskip -1em

\begin{lstlisting}[style=CStyle]
  #include "core.h"
  
  #define bswap_16(x)  (((x & 0xff00) >> 8) | ((x & 0x00ff) << 8))

  // Accel wrapper registers mapping
  #define ACC_SRC_IP   (*((volatile unsigned int  *)(IO_EXT_BASE + 0x00)))
  #define ACC_FW_MATCH (*((volatile unsigned char *)(IO_EXT_BASE + 0x04)))
  
  static inline void slot_rx_packet(struct Desc* desc)
  {
    unsigned short eth_type = *((unsigned short *) (desc->data + 12));
    unsigned int   src_ip   = *((unsigned int   *) (desc->data + 14 + 12));
  
    // check eth type
    if (eth_type == bswap_16(0x0800))
    {
      // start Firewall IP check
      ACC_SRC_IP = src_ip;
      if (ACC_FW_MATCH)
      {
        goto drop;
      }
      else
      {
        desc->port ^= 1;
        pkt_send(desc);
        return;
      }
    }
  
  drop: //Non IPV4 or in firewall list
    desc->len = 0;
    pkt_send(desc);
  }
  
  int main(void)
  {
    // Initializing LB and RPU interconnet
    init_hdr_slots(16, 0x804000, 128);
    init_slots(16, 0x0C000A, 16384);
  
    // Enable only Evict and Poke Interrupts
  	set_masks(0x30); 
  
    while (1)
    {
      // check for new packets
      if (in_pkt_ready())
      {
        struct Desc desc;
        // read descriptor
        read_in_pkt(&desc);
        slot_rx_packet(&desc);
      }
    }
  
    return 1;
  }
\end{lstlisting}

\twocolumn
\section{Artifact Appendix}
\label{app:artifact}

\subsection{Abstract}
Rosebud is a framework to simplify development, debugging and performance tuning
for FPGA-accelerated middleboxes. We have made all of the hardware designs
(Verilog) and software (RISC-V C) open source on GitHub. The RISC-V toolchain
and used python libraries are open-source. Building and programming the FPGA
image requires Xilinx's Vivado toolchain, however building the image take a
long time and requires licenses from Xilinx, so we provide the bitstreams we
used to run the experiments in the paper in the GitHub.  A Makefile to
generate FPGA bitstreams are provided, and also scripts to generate the
host-side and RISC-V binaries.

The primary experiments in this work are evaluating the performance of Rosebud.
Running these experiments requires running two Xilinx FPGA cards. The GitHub
includes experiment scripts to evaluate Rosebud's forwarding throughput and
latency for different packet sizes. It also includes scripts, RISC-V software,
and FPGA images to run the performance tests for the case studies of the Rosebud
framework, including a firewall and an Intrusion Detection System (IDS). We
include scripts to generate the packet traces for these experiments in GitHub.
For each of the experiments we include a README that describes how to
parameterize the Rosebud runtime environment for each data point in the
experiments by passing parameters to a Makefile. We also include in the GitHub
the software simulation framework that can be used to test the Verilog and
RISC-V C for functional correctness.

\vspace {-5pt}
\subsection{Artifact Checklist (Meta-Information)}

{\small
\begin{itemize}
  \item {\bf Run-time environment: } Linux (tested on Ubuntu and Arch). 
  \item {\bf Execution: } Load the Xilinx FPGA bitstreams running on two FPGA cards. Run experiment scripts.
  \item {\bf Metrics: } Throughput (i.e., packets per second), latency.
  \item {\bf Output: } We output the metrics for both the performance tests of
  the framework, as well as the case studies applications. 
  \item {\bf Experiments: } We have provided scripts for the experiments. 
  We expect minor variations from the results in the paper. 
  \item {\bf How much disk space required (approximately)?: } The tests do not
  require considerable disk space, and 100~GBs for Vivado. 
  \item {\bf How much time is needed to prepare workflow?: } An
  hour for installing open source tools, and a few for Vivado. 
  \item {\bf How much time is needed to complete experiments?: } A few hours.
  \item {\bf Workflow framework used?: } We used Makefiles and bash scripts.
  \item {\bf Archived (provide DOI)?: } \url{https://doi.org/10.5281/zenodo.7651655}
\end{itemize}
}

\vspace {-5pt}
\subsection{Description}

\vspace {-2pt}
\subsubsection{How to Access}


\begin{itemize}
    \setlength{\itemsep}{0pt}
    \setlength{\parskip}{0pt}
    \setlength{\parsep}{0pt}
    \item {\bf Publicly available: } \url{https://github.com/ucsdsysnet/Rosebud}
    \item {\bf Code licenses:} MIT License
\end{itemize}

\vspace {-5pt}
\subsubsection{Hardware Dependencies}
\begin{itemize}
    \setlength{\itemsep}{0pt}
    \setlength{\parskip}{0pt}
    \setlength{\parsep}{0pt}
    \item Xilinx Virtex UltraScale+ FPGA VCU1525 (x2)
    \item One or two machines hosting the FPGAs.
\end{itemize}

\vspace {-5pt}
\subsubsection{Software Dependencies}
\begin{itemize}
    \setlength{\itemsep}{0pt}
    \setlength{\parskip}{0pt}
    \setlength{\parsep}{0pt}
    \item Xilinx Vivado 2022.2.1 and its drivers
    \item \texttt{riscv-gcc}
    \item Python3 alongside scapy, pyelftools, dpkt, idstools
    \item \texttt{tcpdump} and \texttt{tcpreplay}
\end{itemize}

\newpage

\subsection{Installation}
Steps to install required open-source software and Vivado drivers are mentioned
in the repository.

\begin{itemize}
    \item {\bf Compilation: } We have provided Makefiles for building the
    host-side binaries, driver, and RISC-V firmware. 
    \item {\bf Binary: } FPGA images are provided.
\end{itemize}
\vspace {-8pt}
\subsection{Experiment Workflow}
First, the FPGAs need to be programmed with the corresponding image. One FPGA is
the tester FPGA that generates test packets, and one is the FPGA running
benchmarks on the \fname framework, which we call the Design Under Test (DUT) FPGA. The tester FPGA is programmed with the
Rosebud framework with a 16-RPU design and is mostly used as a high-speed packet generator. The DUT FPGA is programmed with different configurations of the Rosebud framework to benchmark it under a variety of test cases. 

All of the experiments in this paper follow the same general procedure. First you will load the images on the FPGAs using a script we provide or Vivado. For instance, to load the image on the
DUT FPGA that we use for most of the experiments, go to the directory \directory{host\_utils/runtime} and run:
\begin{lstlisting} [style=BashStyle]
$ ./loadbit.sh prog ../../bitfiles/VCU1525_16RPU_Firewall.bit
\end{lstlisting}

If there are multiple FPGAs on the same machine, you can pass the JTAG index for
the other FPGAs to this script, if this script fails you can use Vivado as a backup method to load the image.

Next, the Corundum Linux driver needs to be built and loaded. Go 
to \directory{host\_utils/driver/mqnic} and do:
\begin{lstlisting} [style=BashStyle]
$ make; sudo modprobe ptp; sudo insmod mqnic.ko
\end{lstlisting}
Then reset the FPGAs after loading the driver,
which can be done by running \bash{\$ make reset\_all} from
\directory{host\_utils/runtime}. If after programming,
PCIe enumeration fails and host cannot see an FPGA (missing in output of 
\bash{make reset\_all}), a system reboot and then reload
of the Corundum driver is required (no need to reprogram). 

To run code on the RISC-V cores on the FPGA, the corresponding binaries should be
generated and loaded to the cores, alongside configuring the system settings
(i.e., receiving RPUs). Finally, the host side profiling utility
is run to measure the throughput of the system. These steps are scripted, e.g., 
in the \bash{make do} command. All evaluation scripts are
located in \directory{host\_utils/runtime}, and trace generation and injection
scripts location are mentioned per each experiment.  If both FPGAs are installed
on the same machine, they can be addressed by their device names, e.g.,
\filename{mqnic0} and \filename{mqnic1}. 

\vspace {-4pt}
\subsection{Evaluation and Expected Results}
In this evaluation, we assume both FPGAs are on the same
machine, where the \filename{mqnic0} device is the DUT FPGA and \filename{mqnic1}
is the tester FPGA. For all experiments, we cross connect the 2 FPGAs with two
100G cables to enable 200~Gbps throughput between them, with the exception of
second step of the latency experiments. 
Using the previous section commands, the tester and DUT FPGAs must be programmed with the 16~RPU packet gen bit file 
(\filename{VCU1525\_16RPU\_Pktgen.bit}), and the Firewall accelerator bit file
(\filename{VCU1525\_16RPU\_Firewall.bit}) respectively. These images are sufficient to carry out all of the experiments except for the Pigasus case
study which requires a separate DUT image.\\

{\bf Packet forwarding throughput (Figure 7):}\\
In two separate shells, go to \directory{host\_utils/runtime} and do:
\begin{lstlisting} [style=BashStyle]
1$ make do TEST=basic_fw RECV=0xffff DEV=mqnic0
2$ make do TEST=basic_pkt_gen RECV=0x0000 DEV=mqnic1 PKT_SIZE=1024
\end{lstlisting}
The first command loads the packet forwarding firmware on the DUT FPGA, and the
second command loads the basic packet generator that generates same size packets
on the tester FPGA. For the packet generator FPGA we set the RPUs with incoming
traffic to none (set RECV flag to 0), as we are only generating packets. 

Now wait for the packets to flow for a minute to get a good average, and stop
the process on the tester FPGA using \emph{Ctrl+C}. The last print of the
status table is the average values, and in the ``RX bytes'' field you can read the
aggregate bytes per second for the physical and virtual Ethernet interfaces,
which, on the DUT FPGA, shows how much data could be absorbed and processed. 
To test different packet sizes, you can stop both commands, and rerun them by
changing the packet size argument for the tester FPGA. This is not necessary, but for more
consistent results you can run \bashs{\$ make reset\_all} to reset the FPGAs before each
test.

For the 8~RPU performance results, you can use the same 16~RPU setup and
disable half of the clusters to achieve the same results. 
On the DUT FPGA side, run the following command:
\begin{lstlisting} [style=BashStyle]
1$ make do TEST=basic_fw RECV=0x00ff ENABLE=0x00ff DEV=mqnic0
\end{lstlisting}
On the tester side repeat the same process for different packet sizes. For 
100~Gbps results, you can update the C code to use single port.

\vspace{4pt}
{\bf Packet forwarding latency (Figure 7):}\\
To compute the forwarding latency, we timestamp the packets before sending them
from the tester FPGA, and after they arrive back after getting forwarded in the DUT
FPGA (timers in all RPUs are synced). This values are periodically sent to the
host. There is a second step that instead of using the DUT FPGA, we
cross-connect ports of the tester FPGA to measure the base latency and deduct it from the
latency measured in the first step.

For the first step, we load the forwarding code on the DUT FPGA:
\begin{lstlisting} [style=BashStyle]
1$ make do TEST=basic_fw RECV=0xaaaa DEV=mqnic0
\end{lstlisting}

We enabled half of the cores for receiving packets to be consistent with
the single FPGA test in the next step, where half RPUs generate packets and 
half receive and report the results to the host.
Throughout this experiment, we can keep the DUT FPGA running.

For the tester FPGA, from \directory{host\_utils/runtime} directory, run:
\bash{2\$ ./run\_latency.sh mqnic1}

This script will loop through different packet sizes, for both low-load and maximum-load
scenarios. It uses \texttt{tcpdump} to capture the latency samples that are sent to the
host. After the script run is finished, you can use this command to extract latency values from the 
received pcaps and calculate the average latency per packet size and in each of
the load levels: \bash{2\$ sudo ./latency\_data\_extractor.sh}

For the second step of this experiment, we rewire the system to make it single
tester FPGA loopback. Now we run the
test script with a minor modification: \bash{2\$ ./run\_latency.sh mqnic1 1l}
This saves the result in a different directory. We can rerun the
\bashs{latency\_data\_extractor.sh} script as before to get the average values. The
final deduction per point is done manually.

\vspace{4pt}
{\bf Firewall case study (number reported in Sec 7.2)}\\
To regenerate the tester packet trace, you should run \bash{\$ make gen} in\\
\directory{fpga\_src/accel/ip\_matcher/python}. This will create a trace based on the firewall
blacklist rules. In this trace, there are 1050 packets which are based on the blacklist, and 4
safe packets.

For the DUT FPGA, from \directory{host\_utils/runtime} run:
\begin{lstlisting} [style=BashStyle]
1$ make do DEST_DIR=../../fpga_src/accel/ip_matcher/c/ TEST=firewall 
  RECV=0xffff DEV=mqnic0
\end{lstlisting}
In another shell for the tester script and in the same directory run:
\begin{lstlisting} [style=BashStyle]
2$ make do TEST=pkt_gen DEV=mqnic1 BLOCK_INTS=3 PKT_SIZE=1024
\end{lstlisting}

The pkt\_gen code makes proper TCP/UDP packets and also can forward traffic from
the host and inject it within the rest of the traffic. We set the \filename{BLOCK\_INTS}
mask so that no traffic is received from the physical Ethernet ports.

Finally, to inject the attack traffic, in another shell go to \\
\directory{fpga\_src/accel/ip\_matcher/python} and run: 
\begin{lstlisting} [style=BashStyle]
3$ make set_mtu DEV=mqnic1; make run DEV=mqnic1
\end{lstlisting}
This will inject the trace at about 5~Gbps. You can keep this packet injector
running, and similar to the forwarding throughput experiment, stop the other two
\bash{make do} shells to observe the RX Bytes on the DUT FPGA, and set a
different packet size for the tester FPGA. If \bash{make reset\_all}
is used in any step, rerun the the injector script.

For application verification, on the tester FPGA you can run:
\begin{lstlisting} [style=BashStyle]
2$ make do TEST=basic_corundum_fw DEV=mqnic1 BLOCK_INTS=3
\end{lstlisting}
The tester FPGA will forward packets received from the host to the DUT
FPGA, and the ratio of TX to RX frames shows the drop rate.

\vspace{4pt}
{\bf Intrusion detection system case study (Figure 8 \& 9)}\\
To generate the traces for this test, from \\
\directory{fpga\_src/accel/pigasus\_sme/python/} run:
\begin{lstlisting} [style=BashStyle]
$ make; mv attack_pcap_* ../pcaps/
\end{lstlisting}

For this experiment we need to reprogram the DUT FPGA, set the RPUs to 8 and
load the RISC-V code. To get the HW reordering
results, run this from \directory{host\_utils/runtime}:
\begin{lstlisting} [style=BashStyle]
1$ ./loadbit.sh prog ../../bitfiles/VCU1525_8RPU_Pigasus_RR_LB.bit
1$ make reset_all
1$ make do DEST_DIR=../../fpga_src/accel/pigasus_sme/c/ TEST=pigasus2 
  RECV=0xffff DEV=mqnic0 RPUS=8
\end{lstlisting}

For the tester FPGA, in another shell run: 
\begin{lstlisting} [style=BashStyle]
2$ make do TEST=pkt_gen DEV=mqnic1 BLOCK_INTS=3 PKT_SIZE=1024
\end{lstlisting}

Finally, for packet injection, from a third shell and from\\
\directory{fpga\_src/accel/pigasus\_sme/pcaps/} directory run:
\begin{lstlisting} [style=BashStyle]
3$ make set_mtu DEV=mqnic1; make attack DEV=mqnic1 SIZE=1024
\end{lstlisting}
This script will adjust the attack rate to be close to one percent.

Similar to the forwarding throughput experiment, you can pause the 
\bash{make do} runs and check the RX bytes on the DUT FPGA.

For SW-based reordering, change the image and firmware and
repeat the same testing process. From \directory{host\_utils/runtime} run:
\begin{lstlisting} [style=BashStyle]
1$ ./loadbit.sh prog ../../bitfiles/VCU1525_8RPU_Pigasus_Hash_LB.bit
1$ make reset_all
1$ make do DEST_DIR=../../fpga_src/accel/pigasus_sme/c/ TEST=pigasus 
  RECV=0xffff DEV=mqnic0 RPUS=8
\end{lstlisting}

The matched packets in this case study are sent to the host and can be seen with
\texttt{tcp\_dump}. Figure 9 results are extracted from the previous
experiments in this subsection, by reversing the frame
rate output to get the average cycles per packet.

Note that experiments can be customized by changing the packet size or
attack traffic that is generated by the Python scripts. 



\clearpage
\bibliographystyle{ACM-Reference-Format}
\balance
\bibliography{confs_long,master}

\end{document}